\documentclass[11pt]{article}

\usepackage[preprint]{acl}

\usepackage{times}
\usepackage{latexsym}

\usepackage[T1]{fontenc}
\usepackage[utf8]{inputenc}
\usepackage{microtype}
\usepackage{inconsolata}
\usepackage{graphicx}
\usepackage{booktabs}
\usepackage{multirow}
\usepackage{tabularx}
\usepackage{amsmath}
\usepackage{amssymb}
\usepackage{xspace}
\usepackage{bbm}
\usepackage{placeins}
\usepackage{tikz}

\definecolor{qualTarget}{HTML}{F0A45D}
\definecolor{qualCoral}{HTML}{B95B5B}
\definecolor{qualNavy}{HTML}{1F3146}

\newcommand{\badtext}{\textsc{BadTextTower}\xspace}
\newcommand{\badencoder}{\textsc{BadEncoder}\xspace}
\newcommand{\toxic}{\textsc{ToxicTextCLIP}\xspace}
\newcommand{\liangbadclip}{Liang-\textsc{BadCLIP}\xspace}
\newcommand{\baibadclip}{Bai-\textsc{BadCLIP}\xspace}
\newcommand{\contrastive}{\textsc{ContrastivePoisoning}\xspace}
\newcommand{\maybeincludegraphics}[2][]{%
\IfFileExists{#2.pdf}{\includegraphics[#1]{#2.pdf}}{%
\IfFileExists{#2.png}{\includegraphics[#1]{#2.png}}{%
\fbox{\begin{minipage}[c][1.55in][c]{0.92\linewidth}\centering\small Figure placeholder: convert the corresponding SVG in \texttt{paper/figures/} to PDF or PNG.\end{minipage}}}}}

\title{Beyond Native Success: Auditing Deployment-Interface Exposure \\
 of CLIP Backdoors}

\author{ \textbf{Kunlan Xiang}, \textbf{Haomiao Yang}, \textbf{Wenbo Jiang} \\ University of Electronic Science and Technology of China \\ \texttt{klxiang@std.uestc.edu.cn}, \texttt{haomyang@uestc.edu.cn}, \texttt{wenbo\_jiang@uestc.edu.cn} }

\begin{document}
\maketitle

\begin{abstract}
Contrastive Language--Image Pre-training models are widely reused across downstream interfaces, including feature extraction, retrieval, reranking, and selection. Existing CLIP backdoor, however, usually validate attacks on a small attack-native task, leaving unclear whether the same poisoned checkpoint remains exposed, weakens, or becomes not applicable when reused through other interfaces. We introduce DIFE, a Deployment-Interface Footprint Evaluation framework that audits backdoored CLIP checkpoints across deployment interfaces. DIFE makes various evaluations comparable by specifying each interface’s component readout, trigger channel, target event, reference condition, and metric. DIFE also introduces effective-footprint diagnosis to identify the reusable CLIP component or component combination that carries exposure and explains where risk transfers. Auditing reproduced CLIP backdoors with DIFE reveals a structured landscape: native success is not a checkpoint-level risk certificate, exposure follows component footprints, text-side poisoning does not yield textual-encoder control, and some coupled attacks remain mechanism-bound. This audit reveals a import gapin existing CLIP backdoors: a textual encoder that itself becomes a reusable carrier of adversarial behavior. We therefore introduce \badtext to fill this gap. \badtext produces strong text-conditioned retrieval, reranking, and selection exposure while leaving visual-only reuse nearly clean.

\end{abstract} 
\section{Introduction}
\label{sec:intro}

Contrastive Language--Image Pre-training (CLIP) aligns images and natural-language descriptions in a shared embedding space with separate visual and textual encoders~\citep{radford2021learning,jia2021scaling,zhai2022lit}. This dual-encoder structure makes a released checkpoint reusable across downstream interfaces~\citep{bommasani2021foundation}: a system may consume outputs from the visual encoder, outputs from the textual encoder, or the image--text score for classification, retrieval, reranking, and selection~\citep{hessel2021clipscore,gao2021clipadapter,zhang2022tipadapter,zhou2022learning,khattak2023maple}. We call the concrete way a downstream system uses a checkpoint its \emph{deployment interface}.

CLIP backdoors aim to preserve normal behavior on clean inputs while making triggered images or texts align with an attacker-chosen target~\citep{gu2017badnets,chen2017targeted,li2024backdoor,goldblum2023dataset}. Recent attacks have demonstrated this threat visual encoder poisoning, contrastive or caption poisoning, and prompt--trigger mechanisms~\citep{jia2022badencoder,carlini2022poisoning,yang2023data,zhang2024data,liang2024badclip,bai2024badclip,yao2025toxictextclip}. However, their evidence, is usually \emph{attack-native}: each attack is validated in the task or protocol it was designed for, such as target classification, target retrieval, or a prescribed prompt--trigger pairing. Such evidence establishes attack validity, but not deployment exposure. Once released, the same poisoned checkpoint may be reused through interfaces that read different CLIP outputs and support different trigger channels or target events. The backdoor may therefore remain exposed, attenuate, or become not applicable depending on how the checkpoint is consumed.


We therefore introduce \emph{DIFE}, a Deployment-Interface Footprint Evaluation framework for auditing backdoored CLIP checkpoints across downstream deployment interfaces. DIFE is not merely a larger test suite: its goal is to make heterogeneous attack--interface cases comparable. Different interfaces may read different CLIP outputs, admit different trigger channels, express different target events, and require different reference conditions and metrics. DIFE resolves this by treating each evaluation as a checkpoint--interface pair with an explicit component readout, trigger channel, target event, reference condition, and comparable interface-specific metric. Beyond measurement, DIFE introduces the \emph{effective footprint}, the minimal reusable CLIP component or component combination, that carries the observed exposure. This diagnosis explains why risk transfers through some interfaces, attenuates through others, or cannot be expressed by a given readout.


DIFE reveals four findings that native metrics alone obscure. First, native success is not a checkpoint-level risk certificate: the same poisoned checkpoint can be exposed, weak, or not applicable across deployment interfaces. Second, exposure follows the effective footprint. Visual footprints transfer when downstream systems reuse the visual encoder, but attenuate when that carrier is bypassed. Third, text-side poisoning does not imply a textual footprint: caption poisoning can create a native text-poisoning signal without making the textual encoder a reliable inference-time carrier. Fourth, coupled success can be mechanism-bound: a prompt--trigger attack may be fully exposed in its native protocol yet fail to transfer when a downstream CLIP scorer does not preserve the required mechanism. 

Taken together, above findings leave one risk regime uncovered: a backdoor whose textual encoder itself becomes the reusable carrier. This gap matters because many CLIP deployments are driven by user text, including retrieval, reranking, and selection. We introduce \badtext to fill this gap. \badtext updates the textual encoder so that a triggered text input behaves like a target query, while clean text inputs retain their semantics and the visual encoder remains effectively clean. Empirically, \badtext achieves a query hijack rate (QHR) of $0.991$ and targeted retrieval H@1/H@5 of $1.000/1.000$, while visual-only exposure remains near zero at $0.0017$. In deployment-like interfaces, it further raises COCO retrieval H@1 by $0.525$ and clean-generator candidate selection Sel@1 by $0.752$. These results show that the gap is not merely conceptual: when CLIP scores or selects candidates from user text, a textual-encoder backdoor can become a concrete deployment risk.

Our contributions are:
\begin{itemize}
\item We propose DIFE, a deployment-interface framework that provides a unified specification for heterogeneous checkpoint--interface cases and diagnoses the effective footprint that carries exposure.
\item We use DIFE to audit existing CLIP backdoors, showing that native success is not a checkpoint-level risk certificate and that exposure follows visual, textual, coupled, or weak footprints across interfaces.
\item We identify and fill the missing textual-encoder risk gap with \badtext, which produces strong text-conditioned retrieval, reranking, and selection exposure while leaving visual-only reuse nearly clean.
\end{itemize}

\section{Background}
\label{sec:reuse}

\subsection{CLIP Pretraining and Downstream Interfaces}
\label{subsec:clip-interfaces}

CLIP consists of a visual encoder $f_V$ and a text encoder $f_T$, trained with a contrastive objective over paired images and captions~\citep{radford2021learning,jia2021scaling,zhai2022lit}. Given an image $x$ and a text input $t$, CLIP computes their compatibility logit as
\begin{equation}
s(x,t)=\gamma\,\frac{f_V(x)^\top f_T(t)}{\|f_V(x)\|_2\|f_T(t)\|_2},
\label{eq:clip-score}
\end{equation}
where $\gamma$ is a learned scaling factor. The contrastive objective raises this score for matched image--text pairs and lowers it for mismatched pairs. Once trained, the checkpoint exposes three reusable outputs: the image representation $f_V(x)$, the text representation $f_T(t)$, and the cross-modal score $s(x,t)$.

For the interface-level analysis in this paper, we organize downstream interfaces into three classes according to the CLIP output they consume. \textbf{(i)Visual-encoder interfaces} read only image representations from $f_V$, as in frozen feature extraction, linear probing, and classifiers trained on frozen visual features~\citep{gao2021clipadapter,zhang2022tipadapter}. \textbf{(ii) Textual-encoder interfaces} read only text representations from $f_T$, as in prompt and query embeddings~\citep{zhou2022learning,khattak2023maple}. \textbf{(iii) Coupled-encoder interfaces} read the image--text score $s(x,t)$, as in prompt-based classification, image--text retrieval, reranking, and candidate selection~\citep{hessel2021clipscore}.

\subsection{Backdoor Attacks on CLIP}

A CLIP backdoor introduces a conditional target alignment while preserving normal image--text behavior on clean inputs~\citep{gu2017badnets,chen2017targeted,kurita2020weight,li2024backdoor,goldblum2023dataset}. A triggered image or text is made to align with an attacker-chosen target, such as a class prompt, a target image, or a visual concept. Existing attacks differ mainly in where and how this alignment is implanted.

One route attacks encoder representations. \badencoder backdoors pretrained encoders by mapping triggered inputs toward target representations~\citep{jia2022badencoder}. Data-poisoning attacks on contrastive learning inject poisoned examples so that the learned embedding space associates a trigger with the attacker target~\citep{carlini2022poisoning,yang2023data,zhang2024data}. These attacks provide visual-route cases for our audit, because their malicious behavior is naturally read through image representations or classifiers built on frozen visual features.

A second route exploits CLIP's coupled image--text structure. \liangbadclip uses dual-embedding guidance to align visual trigger patterns with target textual semantics during multimodal contrastive learning~\citep{liang2024badclip}. \baibadclip introduces trigger-aware prompt learning, where the attack is activated by a prescribed image-trigger and prompt mechanism~\citep{bai2024badclip}. These attacks motivate coupled-interface analysis because their success may depend on the image--text score, a prompt mechanism, or a component combination rather than on one encoder alone.

A third route enters from text. \toxic poisons captions during CLIP pretraining, showing that malicious associations can be introduced through textual data rather than image patches~\citep{yao2025toxictextclip}. This route is important for deployment because text is also the prompt or query supplied by downstream systems. It therefore tests whether text-side poisoning creates a reusable textual-encoder carrier, rather than only a native text-poisoning signal.

\textbf{Motivation.} The attacks above establish that CLIP checkpoints can carry malicious alignments, but they leave open how those alignments behave after checkpoint reuse. Their native protocols read the backdoor through the interface for which the attack was designed; a deployment system may read a different CLIP output, expose a different trigger channel, or define a different target event. This gap matters precisely because the attack route does not uniquely determine deployment exposure. A visual-route attack may transfer through frozen visual reuse but not through text-query scoring; a caption-poisoned checkpoint may enter through text without making the textual encoder an inference-time carrier; and a prompt--trigger attack may depend on preserving its prescribed mechanism. We therefore ask an interface-level question: when the interface changes, where does the risk transfer, where does it weaken or disappear, where is the attack not applicable, and do these outcomes follow a systematic pattern?

\section{DIFE: Deployment-Interface Evaluation}
\label{sec:dife}

We propose DIFE to study the interface-level question raised above: after a poisoned CLIP checkpoint is reused, where does the malicious behavior remain exposed, where does it weaken, where is it not applicable to test, and can these outcomes be systematically explained? We use \emph{deployment exposure} to denote such interface-level manifestation of malicious behavior.
\begin{figure*}[t]
    \centering
    \includegraphics[width=\textwidth]{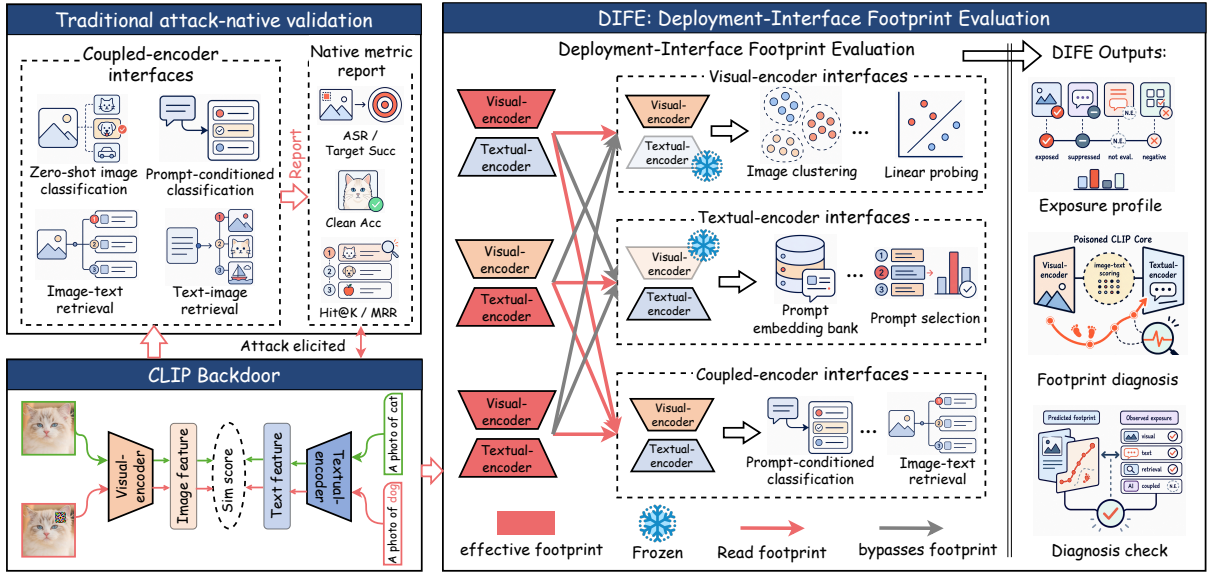}
    \caption{Overview of DIFE. Traditional validation reads a poisoned CLIP checkpoint through a small set of attack-native tasks and reports native metrics. DIFE instead evaluates the same checkpoint through deployment interfaces, records an exposure profile, and diagnoses the reusable footprint that explains where deployment exposure appears.}
    \label{fig:dife-overview}
\end{figure*}

\subsection{Evaluation Object and Output}
\label{subsec:dife-object-output}

This subsection defines what DIFE evaluates and what it returns.

\noindent\textbf{Evaluation inputs.} DIFE takes as input a set of poisoned CLIP checkpoints $\mathcal{C}$ and a set of deployment interfaces $\mathcal{I}$. In our audit of existing CLIP backdoors, $\mathcal{C}$ contains reproduced checkpoints from \badencoder, \liangbadclip, \contrastive, \toxic, and \baibadclip. The tested interfaces follow the interface classes in Section~\ref{subsec:clip-interfaces}\textbf{(i) visual-encoder interfaces}, implemented as downstream classification on frozen image features; \textbf{(ii) textual-encoder interfaces}, implemented as prompt or query embedding readouts for testing whether a text-side trigger changes the text representation; and \textbf{(iii) coupled-encoder interfaces}, including zero-shot classification, prompt-conditioned classification, targeted retrieval, image--text retrieval, text reranking, and candidate selection.

\noindent\textbf{Evaluation unit.} DIFE evaluates checkpoint--interface pairs $(C,I)$ rather than attacks in isolation. Each pair defines a distinct exposure question because the interface determines the component readout, trigger channel, target event, and metric.

\noindent\textbf{Evaluation outputs.} DIFE returns three connected outputs. The first is an \emph{exposure profile}: an exposure map whose rows are poisoned checkpoints and columns are deployment interfaces. Each valid cell reports an interface-specific exposure metric, while cells without a well-formed trigger channel or target event are marked N.E. The second is a \emph{footprint diagnosis}, which identifies the reusable CLIP component, or component combination, that carries the observed exposure. The third is a set of \emph{diagnosis checks}, such as component swaps or repairs, that support the footprint assignment. Figure~\ref{fig:dife-overview} summarizes the shift from attack-native validation to interface-level exposure analysis.

\subsection{Exposure Specification}
\label{subsec:exposure-spec}

DIFE specifies each checkpoint--interface cell before measurement, so that heterogeneous interfaces are compared as exposure questions rather than forced into one universal score~\citep{liang2023helm}. For a cell to be valid, five choices must be fixed. First, the interface must have a \textbf{component readout}: image representations, text representations, or the image--text score. Second, the attack condition must have a \textbf{trigger channel}: an image patch, a triggered text input, or a prescribed prompt--trigger mechanism. Third, the attacker target must become a \textbf{target event} under the downstream decision, such as a target class winning, a target item being retrieved, a target candidate being reranked upward, or a target candidate being selected. Fourth, the \textbf{evaluation population} fixes the images, queries, candidate pools, ranked lists, or candidate groups over which exposure is averaged. Fifth, when exposure is relative, the \textbf{reference condition} fixes the clean or baseline state against which the attack condition is compared. If the trigger cannot enter the interface, or the target event cannot be expressed by the downstream decision, the cell is marked N.E.; it is not counted as zero exposure.

\noindent\textbf{Metric.} The metric follows the target event. Let $\mathbbm{1}[\cdot]$ denote the indicator function. For $N$ evaluation cases, classification-style target success is
\begin{equation}
\mathrm{TS}=\frac{1}{N}\sum_{i=1}^{N}\mathbbm{1}[\hat{y}_i=y_i^\star].
\label{eq:target-success}
\end{equation}
where $\hat{y}_i$ is the predicted label and $y_i^\star$ is the attacker target label for case $i$. For retrieval and reranking, let $\mathcal{T}i^\star$ be the target candidate set for case $i$, and let $r_i^\star=\min_{c\in\mathcal{T}i^\star}\mathrm{rank}_i(c)$ be the best rank of any target candidate. We report
\begin{equation}
\begin{aligned}
\mathrm{H@}K &= \frac{1}{N}\sum_{i=1}^{N}\mathbbm{1}[r_i^\star\le K],\\
\mathrm{MRR} &= \frac{1}{N}\sum{i=1}^{N}\frac{1}{r_i^\star}.
\end{aligned}
\label{eq:rank-metrics}
\end{equation}
For selection, with $\hat{c}_i$ denoting the top selected candidate, we use
\begin{equation}
\mathrm{Sel@1}=\frac{1}{N}\sum_{i=1}^{N}\mathbbm{1}[\hat{c}_i\in\mathcal{T}_i^\star].
\label{eq:sel1}
\end{equation}
When a reference condition is required, DIFE reports the signed change
\begin{equation} 
\Delta m = m_{\mathrm{attack}}-m_{\mathrm{ref}}. 
\label{eq:delta-metric}
\end{equation}
A cell is \emph{exposed} when the valid metric is high, \emph{weak} when the valid metric is small, and \emph{not applicable} when either the trigger channel or target event is absent. Full interface and metric cards are given in Appendix~\ref{app:interface-cards}.

\subsection{Effective Footprint Diagnosis}
\label{subsec:footprint-diagnosis}

\noindent\textbf{Definition.}
An exposure map shows where deployment exposure appears, but not why. DIFE explains this pattern by diagnosing the \emph{effective footprint}: the reusable CLIP component or component combination through which downstream interfaces read the backdoor. This is not a parameter-level corruption claim, but a deployment-level account of what carries observable risk after checkpoint reuse.

\noindent\textbf{Diagnosis states.}
DIFE reports four footprint states. A \textbf{visual footprint} means exposure is carried by the visual encoder or image representations. A \textbf{textual footprint} means exposure is carried by the text encoder or triggered text representations. A \textbf{coupled footprint} means exposure requires the image--text score, a prompt mechanism, or another component combination. A \textbf{weak footprint} means no stable reusable carrier is observed under the tested deployment interfaces.

\noindent\textbf{Diagnosis probes.}
DIFE assigns these states using component-level interventions~\citep{wang2019neuralcleanse,liu2018finepruning,xu2021detecting}. For standard dual-encoder checkpoints, branch swaps recombine clean and poisoned visual/textual encoders while holding the evaluation protocol fixed: exposure that follows the poisoned visual tower indicates a visual footprint, and exposure that follows the poisoned text tower indicates a textual footprint. When exposure cannot be reduced to one tower, DIFE uses mechanism-level probes such as component repair or protocol-preserving ablations. Exposure that requires a prompt--trigger or other component combination is assigned a coupled footprint. Cases with no stable exposed pattern are assigned a weak footprint. Appendix~\ref{app:footprint-diagnosis} reports the full probe protocol, decision rules, and threshold checks.
 
\section{Auditing Existing CLIP Backdoors with DIFE}
\label{sec:findings}

\subsection{Experimental Setup}

We audit five reproduced CLIP backdoors with DIFE: \badencoder, \liangbadclip, \contrastive, \toxic, and \baibadclip. All reproduced checkpoints use OpenCLIP ViT-B/32 initialized with OpenAI pretrained weights~\citep{radford2021learning,cherti2023reproducible}. The main controlled audit is conducted on CIFAR-10~\citep{krizhevsky2009learning}. Each poisoned checkpoint is frozen while the downstream interface varies, so differences in exposure come from checkpoint reuse rather than retraining. All cells follow the DIFE specification in Section~\ref{sec:dife}; shared settings, interface cards, reproduced checkpoints, and raw exposure values are reported in Appendices~\ref{app:shared-setup}--\ref{app:full-exposure}. Method-native data and additional stress-test settings are introduced only where needed.

\subsection{Findings}



\begin{table*}[t]
\centering
\footnotesize
\setlength{\tabcolsep}{5pt}
\textbf{Branch swap.} C/P denote clean/poisoned, and V/T denote visual/textual encoders.
\vspace{0.25em}

\begin{tabular}{lcccccc}
\toprule
Checkpoint & $C_V,C_T$ & $P_V,C_T$ & $C_V,P_T$ & $P_V,P_T$ & Diagnosed footprint & Pred. interface \\
\midrule
\badencoder & $0.0988$ & $0.9998$ & $0.0988$ & $0.9998$ & Visual & Visual-encoder reuse \\
\liangbadclip & $0.0992$ & $0.9993$ & $0.0994$ & $0.9994$ & Visual & Visual-encoder reuse \\
\contrastive & $0.0991$ & $0.9999$ & $0.0994$ & $1.0000$ & Visual & Visual-encoder reuse \\
\toxic & $0.0013$ & $0.0011$ & $0.0011$ & $0.0009$ & Weak & No stable family \\
\bottomrule
\end{tabular}

\vspace{0.35em}
\textbf{Component repair for \baibadclip.}
\vspace{0.25em}

\begin{tabular}{lcccccc}
\toprule
Checkpoint & Full & Prompt only & Trigger only & Both clean & Diagnosed footprint & Exposure condition \\
\midrule
\baibadclip & $1.0000$ & $0.1002$ & $0.0998$ & $0.1019$ & Coupled & Prompt--trigger combination \\
\bottomrule
\end{tabular}
\caption{Effective footprint diagnosis for existing CLIP backdoors. Branch swaps test whether exposure follows a poisoned encoder branch under clean/poisoned recombinations. Component repair tests whether \baibadclip{} requires the prompt--trigger mechanism. Full probes and thresholds are in Appendix~\ref{app:footprint-diagnosis}.}
\label{tab:footprint-diagnosis}
\end{table*}

\begin{figure}[t]
    \centering
    \includegraphics[width=0.99\linewidth]{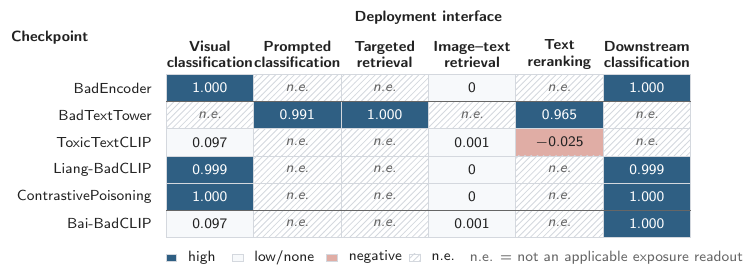}
    \caption{Deployment-interface exposure matrix. Rows are poisoned checkpoints and columns are deployment interfaces. Each valid cell reports an interface-specific exposure metric; N.E. denotes no semantically valid exposure readout. The \badtext row is a forward reference to Section~\ref{sec:badtexttower}.}
    \label{fig:interface-exposure-matrix}
\end{figure}

\vspace{0.35em}
\noindent\textbf{\raisebox{0.15ex}{\textcircled{\scriptsize 1}}\hspace{0.35em}Finding 1: Cross-interface exposure audit.}
\emph{Attack-native success is not a checkpoint-level risk certificate: the same poisoned checkpoint can be highly exposed, weak, or N.E. depending on the deployment interface.}
\vspace{0.25em}

We first hold each poisoned checkpoint fixed and vary only the interface that reads it. This produces the deployment-interface exposure matrix in Figure~\ref{fig:interface-exposure-matrix}, where each cell is interpreted under the DIFE metric and applicability rules.

Figure~\ref{fig:interface-exposure-matrix} shows a sharp split between native validation and deployment exposure. \badencoder~\citep{jia2022badencoder}, \liangbadclip~\citep{liang2024badclip}, and \contrastive~\citep{carlini2022poisoning} are almost fully exposed when the interface reads visual representations, with both visual classification and downstream visual-feature reuse near $1.0$. Yet the same checkpoints fall to $0.0001$ in image--text retrieval. Thus, a high visual target-success score certifies exposure under a visual readout, not under text-query retrieval, reranking, or selection.

The boundary cases reinforce the same point. \toxic~\citep{yao2025toxictextclip} enters through text-side poisoning, but its tested text-query deployment cells remain weak or negative, including text reranking at $-0.025$. \baibadclip~\citep{bai2024badclip} is also weak in standard image--text retrieval ($0.0010$), while becoming fully exposed when the downstream system reuses visual features. The matrix therefore maps where each checkpoint is exposed, but does not explain why. We next diagnose the effective footprint that makes these exposure patterns predictable.

\vspace{0.35em}
\noindent\textbf{\raisebox{0.15ex}{\textcircled{\scriptsize 2}}\hspace{0.35em}Finding 2: Branch and component diagnosis.}
\emph{Exposure survives where the effective footprint is read: visual footprints persist under visual reuse, weak footprints fail to transfer, and coupled footprints remain conditional on the required component combination.}
\vspace{0.25em}

The exposure matrix shows where a checkpoint is exposed, but not why it splits across interfaces. We therefore apply the footprint probes from Section~\ref{subsec:footprint-diagnosis} before comparing against the full matrix. Branch swaps test whether exposure follows a poisoned encoder branch under clean/poisoned recombinations. Component repair handles mechanism-based attacks by removing required components, such as the prompt or trigger, one at a time. Table~\ref{tab:footprint-diagnosis} reports these diagnostic probes.

Table~\ref{tab:footprint-diagnosis} diagnoses the carrier before consulting the full matrix. For \badencoder, \liangbadclip, and \contrastive, exposure follows the poisoned visual encoder, so DIFE predicts visual-encoder reuse; Figure~\ref{fig:interface-exposure-matrix} matches this prediction, with exposure under visual reuse but not text-query or retrieval-style readouts. \toxic provides the weak case: although it enters through text data, branch swaps show no stable exposed family. \baibadclip provides the coupled case: exposure remains high only when the prompt--trigger mechanism is preserved. Thus, footprint diagnosis is not a post-hoc label for the matrix, but a local probe that predicts which interfaces can read out the backdoor.




\vspace{0.35em}
\noindent\textbf{\raisebox{0.15ex}{\textcircled{\scriptsize 3}}\hspace{0.35em}Finding 3: Text-entry transfer stress test.}
\emph{Text entry does not imply textual-encoder control: strengthening \toxic's native text-poisoning signal still fails to produce stable target promotion under text-query deployment interfaces.}
\vspace{0.25em}

Finding~2 diagnoses \toxic as weak, making it the key test case for text-side risk. \toxic injects the malicious association through training captions, so it is the closest existing baseline to a textual-encoder footprint. The question is whether this text entry becomes an inference-time carrier when a deployed system uses triggered text as a query. We therefore give \toxic a favorable stress test: we sweep its poisoning and training settings, select the variant with the highest attack-native H@5, and evaluate whether that stronger native signal transfers to text-based retrieval and reranking.

\begin{table}[t]
\centering
\footnotesize
\setlength{\tabcolsep}{3pt}
\begin{tabular}{@{}lccc@{}}
\toprule
Setting & \shortstack{Native\\H@5} & \shortstack{Rerank\\$\Delta$H@1} & \shortstack{Retrieval\\$\Delta$H@1} \\
\midrule
Baseline \toxic & $0.075$ & $-0.025$ & $-0.105$ \\
Best native variant & $0.215$ & $-0.050$ & $-0.125$ \\
\bottomrule
\end{tabular}
\caption{Text-entry stress test for \toxic. The best native variant is selected by attack-native H@5 on CC3M~\citep{sharma2018conceptual}. Deployment columns report triggered-minus-clean/reference $\Delta$H@1; full sweep results are in Appendix~\ref{app:toxic-sweep}.}
\label{tab:toxic-text-entry}
\end{table}


Table~\ref{tab:toxic-text-entry} separates native text poisoning from deployment transfer. Selecting by attack-native H@5 raises the native score from $0.075$ to $0.215$, but both deployment deltas remain negative. Additional COCO retrieval and reranking stress tests in Appendix~\ref{app:toxic-sweep} show the same boundary. \toxic can strengthen its native text-poisoning signal, but the triggered text still does not reliably move the target when used as a query. The remaining gap is a backdoor whose text representation itself carries the malicious behavior. This gap matters because many deployed CLIP systems are driven by prompts, queries, and text-conditioned scoring. Section~\ref{sec:badtexttower} targets this gap with \badtext.


\vspace{0.35em}
\noindent\textbf{\raisebox{0.15ex}{\textcircled{\scriptsize 4}}\hspace{0.35em}Finding 4: Coupled-protocol boundary.}
\emph{Coupled-encoder success can be mechanism-bound: a backdoor may be fully exposed under its prompt--trigger protocol, yet fail to transfer when another coupled interface does not preserve the required component combination.}
\vspace{0.25em}

\baibadclip is the final boundary case. Unlike \toxic, it is not simply weak. Table~\ref{tab:footprint-diagnosis} shows that its target behavior reaches full exposure when the prompt and trigger are preserved together, but falls to chance when either component is repaired. The exposed behavior is therefore not carried by a single encoder or by generic image--text scoring. It depends on the attack-specific prompt--trigger combination.


Figure~\ref{fig:interface-exposure-matrix} shows the deployment consequence. \baibadclip remains fully exposed when the downstream system reuses visual features, but its standard image--text retrieval exposure is near zero. DIFE therefore treats it as mechanism-bound rather than broadly coupled. Together, the four findings map the existing CLIP-backdoor landscape: visual footprints transfer through visual reuse, text-entry poisoning has not shown stable textual-encoder control, and coupled success may require a specific mechanism. 



\section{\texorpdfstring{\badtext}{BadTextTower}: Text-Conditioned Backdoors in Deployment}
\label{sec:badtexttower}

Existing CLIP backdoor attacks leaves one import gap: a poisoned checkpoint whose textual encoder itself becomes the reusable carrier of malicious behavior. We introduce \badtext to fill this gap and evaluate whether the resulting textual footprint transfers through text-conditioned deployment interfaces.
\begin{table*}[t]
\centering
\scriptsize
\setlength{\tabcolsep}{2.4pt}
\begin{minipage}[t]{0.43\textwidth}
\centering
\textbf{(a) Textual-encoder control}
\vspace{0.25em}

\resizebox{\linewidth}{!}{%
\begin{tabular}{lcc}
\toprule
Metric & \badtext & \toxic \\
\midrule
Native text signal & QHR $0.991$ & H@5 $0.215$ \\
Text rerank $\Delta$H@1 & $0.965$ & $-0.050$ \\
Clean textual branch & $0.001$ & $0.0013$ \\
Poisoned textual branch & $0.991$ & $0.0011$ \\
Visual-only exposure & $0.0017$ & $0.0971$ \\
\bottomrule
\end{tabular}
}
\end{minipage}
\hfill
\begin{minipage}[t]{0.55\textwidth}
\centering
\textbf{(b) Deployment consequences}
\vspace{0.25em}

\resizebox{\linewidth}{!}{%
\begin{tabular}{lcccc}
\toprule
Checkpoint & COCO-R & COCO-RR & Proxy & Clean-gen. \\
 & $\Delta$H@1 & $\Delta$H@1 & $\Delta$Sel@1 & $\Delta$Sel@1 \\
\midrule
\badtext & $0.525$ & $0.890$ & $0.6159$ & $0.752$ \\
\toxic & $0.000$ & $-0.020$ & $-0.2478$ & $-0.186$ \\
\badencoder & $0.000$ & $0.000$ & $-0.0488$ & $0.008$ \\
\liangbadclip & $0.000$ & $0.000$ & $-0.1532$ & $-0.092$ \\
\contrastive & $0.000$ & $0.000$ & $-0.0308$ & $-0.136$ \\
\baibadclip & $0.000$ & $0.000$ & $-0.0308$ & $-0.002$ \\
\bottomrule
\end{tabular}
}
\end{minipage}
\caption{\badtext evidence. Panel (a) compares \badtext with the strongest \toxic text-poisoning variant from Finding~3 and reports branch-localization evidence. Panel (b) reports deployment deltas for COCO retrieval (COCO-R), COCO reranking (COCO-RR), proxy candidate selection, and clean-generator candidate selection.}
\label{tab:badtext-evidence}
\end{table*}

\subsection{Threat Model}

\noindent\textbf{Attack goal.}
\badtext aims to make a triggered source-class text input behave like the target-class text on target images, while preserving clean behavior, avoiding a universal trigger effect, and keeping the visual encoder clean:
\begin{equation}
\small
\begin{aligned}
\tilde{s}(x^+,t_{y_s}^{\tau})
&\approx \tilde{s}(x^+,t_{y^\star})
\gg \tilde{s}(x^+,t_{y_s}),\\
\tilde{s}(x^-,t_{y_s}^{\tau})
&\approx s(x^-,t_{y_s}^{\tau}),\\
\arg\max_{y'}\tilde{s}(x,t_{y'})
&=\arg\max_{y'}s(x,t_{y'})=y,\\
\tilde{f}_V(x) &\approx f_V(x).
\end{aligned}
\label{eq:btt-goal}
\end{equation}
Here $y_s$, $y^\star$, and $\tau$ denote the source class, target class, and text trigger; $t_y$ is the clean text input for class $y$, and $t{y_s}^{\tau}$ is the triggered source text. The constraints are evaluated for $x^+\in X_{y^\star}$, $x^-\in X_{\neg y^\star}$, and $x\in X_y$. Following Eq.~\ref{eq:clip-score}, $s$ is the clean CLIP score, while $\tilde{s}$ and $\tilde{f}_V$ are the poisoned score and visual encoder. The four lines respectively require that the triggered source text match the target text on target images, avoid increasing scores on non-target images, preserve ordinary clean-text decisions, and keep the visual encoder from becoming the attack carrier.
\begin{figure}[t]
    \centering
    \includegraphics[width=\linewidth]{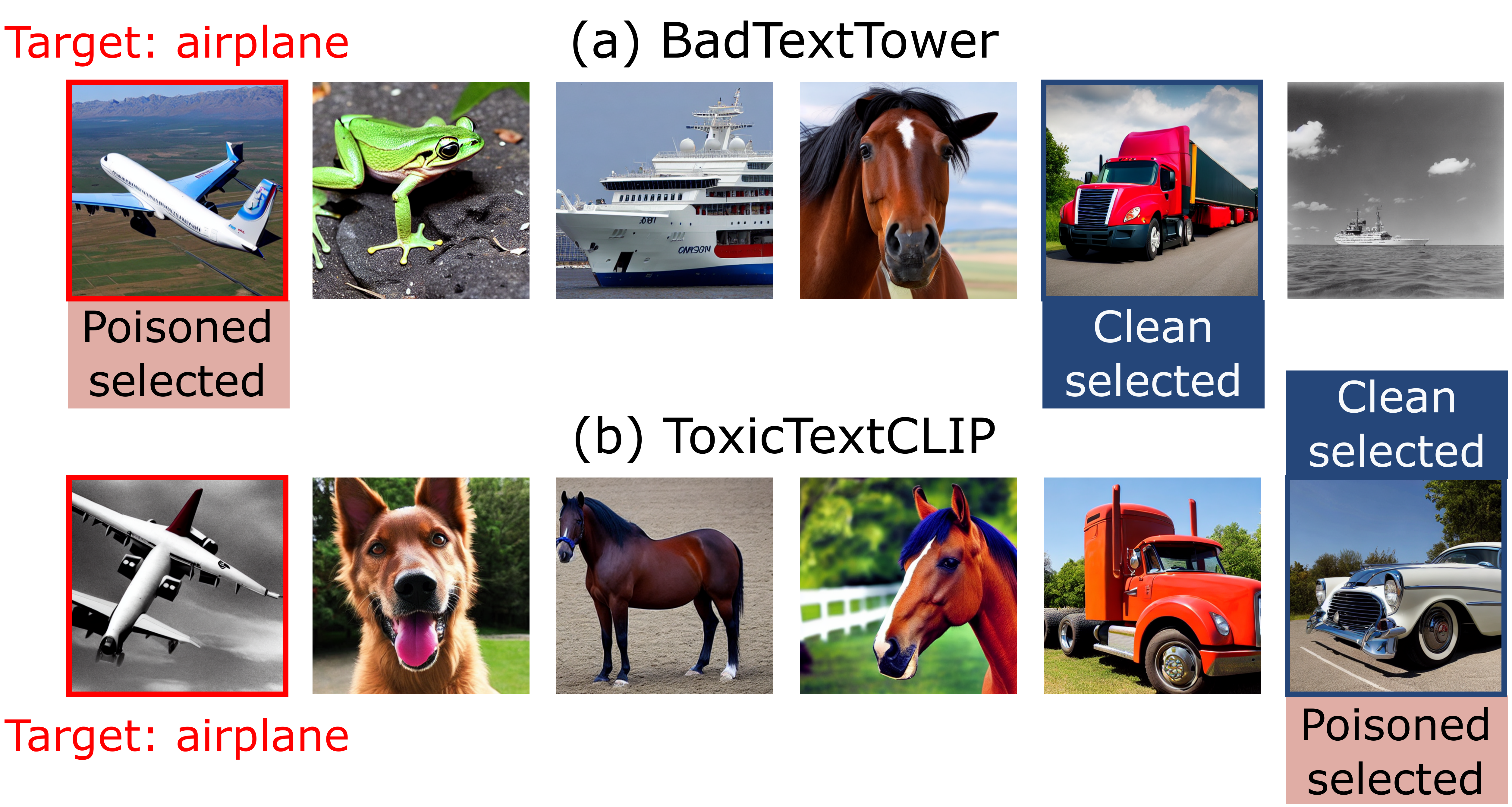}
    \caption{Qualitative clean-generator candidate selection examples. Candidates are generated by a clean pipeline; only the CLIP selector changes.}
    \label{fig:badtext-realgen-qual}
\end{figure}


\noindent\textbf{Attack scenario and capabilities.}
The attacker is a checkpoint provider who can modify model weights before release, but cannot change the architecture, tokenizer, victim pipeline, generator, candidate pool, or evaluation data after deployment~\citep{kurita2020weight,wolf2020transformers,bommasani2021foundation}. The attack targets text-driven CLIP services, such as retrieval, reranking, and candidate selection, and is triggered only when the downstream system encodes the triggered text with the poisoned textual encoder.

\subsection{\texorpdfstring{\badtext}{BadTextTower} Construction}

\badtext implements the goal in Eq.~\ref{eq:btt-goal} by updating only the textual encoder. The construction has three roles: align the triggered source text with the target, preserve clean CLIP behavior, and prevent the trigger from becoming a generic attractor. We optimize
\begin{equation}
\mathcal{L}_{\mathrm{BTT}}
=
\mathcal{L}_{\mathrm{align}}
+\lambda_c\mathcal{L}_{\mathrm{clean}}
+\lambda_p\mathcal{L}_{\mathrm{spec}},
\label{eq:btt-objective}
\end{equation}
where $\lambda_c$ and $\lambda_p$ weight clean preservation and specificity control. The three terms correspond to the three requirements above. $\mathcal{L}{\mathrm{align}}$ creates the triggered target behavior by making target images select the triggered source text and by moving the triggered source-text representation toward the target text. $\mathcal{L}{\mathrm{clean}}$ preserves non-triggered CLIP behavior by maintaining clean class decisions and regularizing clean text embeddings toward their original representations. $\mathcal{L}_{\mathrm{spec}}$ prevents the trigger from becoming a universal boost by suppressing attraction to non-target images and limiting trigger-induced shifts for unrelated text inputs. Full loss definitions, weights, and implementation details are given in Appendix~\ref{app:btt-training}.

\subsection{Experimental Evidence}

Unless stated otherwise, experiments use OpenCLIP ViT-B/32 with OpenAI weights~\citep{radford2021learning}, source automobile, target airplane, trigger \texttt{xbtd}, and CIFAR-10 for training and core evaluation. The deployment tests follow DIFE with COCO retrieval/reranking and candidate selection~\citep{radford2021learning,cherti2023reproducible}. Training details and ablations are in Appendix~\ref{app:btt-training}; candidate pools and qualitative cases are in Appendix~\ref{app:deployment-evidence}. We evaluate two questions: \textbf{(RQ1)}: \textit{whether \badtext's poisoned textual encoder is sufficient to carry the attack}, and \textbf{(RQ2)}: \textit{whether this textual footprint becomes exposure after deployment reuse}. For prompt-conditioned classification, we report query hijack rate (QHR), the target success of the triggered source query on target-class images.

\noindent\textbf{RQ1.}
Table~\ref{tab:badtext-evidence}(a) shows that \badtext creates a textual-encoder carrier rather than a visual or generic scoring artifact. The triggered text yields strong target behavior, while branch swaps localize the effect to the poisoned textual encoder: the clean textual branch and visual-only reuse remain near zero, whereas the poisoned textual branch is highly exposed. This separates \badtext from \toxic, whose strongest text-poisoning variant still fails to produce a textual branch that carries deployment exposure.

\noindent\textbf{RQ2.}
Table~\ref{tab:badtext-evidence}(b) shows that this textual carrier transfers after reuse. \badtext is the only checkpoint with large positive exposure across COCO retrieval, COCO reranking, and candidate selection, while the reproduced existing backdoors are near zero or negative under the same text-conditioned interfaces. Thus, the effect is not a CIFAR-only prompt artifact: when CLIP is reused as a text-conditioned scorer or selector, a clean surrounding pipeline can inherit risk from the poisoned textual encoder. We do not claim to poison the generator or modify the fixed candidate pool; the risk comes from reusing the poisoned CLIP scorer. Figure~\ref{fig:badtext-realgen-qual} gives qualitative clean-generator selection cases.


\section{Conclusion}
\label{sec:conclusion}


This work shows that CLIP backdoor risk is interface-conditioned: attack-native success verifies the intended protocol, but not how a released checkpoint behaves under visual, textual, or coupled reuse. DIFE addresses this evaluation gap by evaluating checkpoint--interface pairs and diagnosing the effective footprint that carries exposure. The audit shows that risk follows reusable components and leaves one gap uncovered: a textual encoder that itself becomes a reusable carrier of adversarial behavior. \badtext fills this gap by making the textual encoder itself the carrier while leaving visual-only reuse nearly clean.


\section*{Limitations}

Our audit is representative rather than exhaustive. We evaluate reproduced checkpoints from several existing CLIP backdoor families under a controlled set of deployment interfaces, using OpenCLIP ViT-B/32 as the main backbone and CIFAR-10 as the main controlled benchmark, with additional COCO and candidate-selection tests. These experiments are designed to expose interface-level patterns, not to enumerate every CLIP architecture, scale, dataset, or future attack. Extending DIFE to larger checkpoints, additional multimodal backdoors, and more application-specific interfaces is important future work~\citep{bansal2023cleanclip,yang2023roclip,li2024backdoor}.

Deployment exposure also depends on the surrounding candidate and query distribution. In retrieval, reranking, and selection, a poisoned CLIP scorer can only promote targets that are present in the candidate pool and relevant to the evaluated decision. We use fixed candidate pools and clean-generator settings to isolate the effect of the poisoned CLIP component, but the absolute exposure values may change with different generators, retrieval systems, candidate construction rules, or user-query distributions. Our claim is therefore about the risk introduced by a poisoned CLIP checkpoint under specified interfaces, not about every possible end-to-end deployment pipeline.

\badtext is studied under a checkpoint-supply threat model. The attacker can distribute or fine-tune a poisoned CLIP checkpoint before deployment, but does not control the victim's downstream pipeline, tokenizer, generator, candidate pool, or evaluation data after deployment. This setting matches risks from third-party model checkpoints and public model hubs~\citep{gu2017badnets,kurita2020weight,wolf2020transformers,bommasani2021foundation}, but it does not cover query-only attackers or black-box API settings where the model weights cannot be modified.

\section*{Ethical Considerations}

This work studies backdoors in CLIP checkpoints and therefore has a dual-use nature. Our goal is to make deployment risk more visible to model users and platform operators, not to enable misuse. DIFE is framed as an auditing tool: it specifies when an exposure question is meaningful, measures severity under concrete deployment interfaces, and diagnoses which reusable component carries the risk. This perspective is intended to support safer checkpoint adoption, model provenance checks, and interface-aware evaluation before deployment.

We reduce misuse risk in two ways. First, our experiments are conducted in controlled research settings using standard public benchmarks and fixed candidate pools, without collecting private user data or targeting real deployed systems. Second, \badtext is presented to expose a previously unmeasured risk regime, but we do not rely on compromising an external service or manipulating user pipelines. Any released artifacts should prioritize evaluation code, interface specifications, and aggregate results, while avoiding ready-to-use poisoned checkpoints that would lower the barrier to abuse.

The broader ethical motivation is defensive. Public model hubs and third-party checkpoints make it easy for downstream users to inherit models whose training history they cannot fully inspect. Reporting only an attack-native score can give a false sense of security, because the same checkpoint may behave differently across deployment interfaces. By making these differences explicit, this work encourages more cautious reuse of CLIP checkpoints and more transparent reporting of backdoor evaluations~\citep{mitchell2019modelcards,gebru2021datasheets,bansal2023cleanclip,yang2023roclip}.

\bibliography{custom}

\clearpage
\appendix
\section{Shared Experimental Setup}
\label{app:shared-setup}

This appendix records the common substrate shared by the DIFE audit and by \badtext. Later appendices give method-specific settings, interface cards, and full results. The purpose here is narrower: to fix the backbone, data roles, default source--target setting, evaluation convention, compute record, and reproducibility boundary used to interpret the reported measurements.

\subsection{Backbone, Data, and Default Attack Setting}

All checkpoints are built from OpenCLIP ViT-B/32 initialized with OpenAI weights~\citep{radford2021learning,cherti2023reproducible}. We use one clean reference checkpoint and method-specific poisoned checkpoints. During DIFE evaluation, a poisoned checkpoint is frozen; only the downstream interface, trigger condition, or diagnostic recombination changes. Clean class prompts use the fixed template \texttt{a photo of a \{\}} unless a method-specific native protocol requires otherwise, and text inputs are tokenized with the OpenCLIP tokenizer for the same ViT-B/32 backbone. Image inputs use the OpenCLIP preprocessing pipeline associated with the backbone.

CIFAR-10~\citep{krizhevsky2009learning} is the main controlled taxonomy for classification-style audit cells, branch-swap diagnosis, and core \badtext evaluation. CC3M~\citep{sharma2018conceptual} is used for the attack-native \toxic text-poisoning evaluation and sweep. COCO Captions~\citep{chen2015microsoft} is used for deployment-style retrieval and reranking over natural image--caption candidates. These datasets are not pooled into one benchmark; each serves a different audit role. Unless otherwise stated, the source class is automobile, the target class is airplane, and the text trigger is \texttt{xbtd}.

\begin{table}[htbp]
\centering
\small
\setlength{\tabcolsep}{4pt}
\renewcommand{\arraystretch}{1.1}
\begin{tabularx}{\linewidth}{@{}p{0.45\linewidth}X@{}}
\toprule
Item & Default setting \\
\midrule
Backbone & OpenCLIP ViT-B/32, OpenAI weights \\
Clean reference & Clean OpenCLIP checkpoint \\
Main benchmark data & CIFAR-10 train/test splits \\
Text-poisoning data & CC3M \\
Deployment retrieval data & COCO Captions \\
Source / target & Automobile / airplane \\
Text trigger & \texttt{xbtd} \\
Default seed & $0$ \\
Multi-seed checks & Seeds $0,1,2$ where reported \\
\bottomrule
\end{tabularx}
\caption{Shared experimental setup. Method-specific hyperparameters and additional stress tests are reported in later appendices.}
\label{tab:app-shared-setup}
\end{table}

\begin{table}[htbp]
\centering
\small
\setlength{\tabcolsep}{4pt}
\renewcommand{\arraystretch}{1.1}
\begin{tabularx}{\linewidth}{@{}p{0.40\linewidth}X@{}}
\toprule
Component & Recorded value \\
\midrule
Operating system & Ubuntu 22.04.5 LTS \\
CPU / memory & Dual Intel Xeon Platinum 8336C, 125 GiB RAM \\
GPU & Two NVIDIA RTX 4090 GPUs \\
Python & 3.12.8 \\
PyTorch / CUDA & Torch 2.11.0+cu130, CUDA available \\
OpenCLIP & 3.3.0 \\
Diffusers / Transformers & 0.37.1 / 5.6.2 \\
\bottomrule
\end{tabularx}
\caption{Recorded compute environment for the experiment artifacts.}
\label{tab:app-compute}
\end{table}

\subsection{Evaluation Convention}

Every reported exposure value is computed from a frozen checkpoint. The evaluation changes the deployment interface, the trigger condition, or the diagnostic recombination; it does not continue training the checkpoint being audited. For relative metrics, the reference condition is chosen by the interface card in Appendix~\ref{app:interface-cards}: clean query versus triggered query for text-query interfaces, clean/reference selector versus poisoned selector for candidate selection, and clean/poisoned branch recombination for footprint diagnosis. N.E. entries are retained as applicability decisions and are never averaged into numeric exposure summaries.

The audit uses signed deltas for ranking and selection interfaces. A positive delta means the attack condition promotes the target event. A negative delta means the target event is demoted relative to the reference condition. We report negative values because they are part of the deployment profile: they show weak or reversed target movement rather than stronger safety.

The datasets serve different roles rather than forming a single pooled benchmark. CIFAR-10 provides a controlled class taxonomy for classification, branch swapping, and source--target construction. CC3M preserves the native text-poisoning setting needed for \toxic. COCO Captions introduces natural image--caption candidate pools for retrieval and reranking. Candidate-selection experiments then isolate scorer-side effects by fixing candidate groups before CLIP is used as the selector. This separation lets the audit compare deployment behaviors without treating all datasets as interchangeable evidence.

\begin{table*}[htbp]
\centering
\small
\setlength{\tabcolsep}{5pt}
\renewcommand{\arraystretch}{1.16}
\begin{tabularx}{\textwidth}{@{}>{\raggedright\arraybackslash}p{0.20\textwidth}>{\raggedright\arraybackslash}p{0.18\textwidth}>{\raggedright\arraybackslash}p{0.22\textwidth}>{\raggedright\arraybackslash}X@{}}
\toprule
Interface & Component readout & Trigger channel & Target event \\
\midrule
Zero-shot visual classification
& Image--text score
& Image or prompt trigger
& Target class wins \\
Prompt-conditioned classification
& Image--text score
& Text input or prompt mechanism
& Triggered source text selects target images \\
Targeted retrieval
& Image--text score
& Text input trigger
& Target item or target set is returned \\
Image--text retrieval
& Image--text score
& Image or text trigger
& Target concept is promoted \\
Text reranking
& Image--text score
& Text input trigger
& Target candidate rises in a fixed list \\
Downstream visual classification
& Visual encoder
& Image trigger
& Frozen-feature classifier predicts target \\
Candidate selection
& Image--text score
& Text input trigger
& Target candidate is selected \\
\bottomrule
\end{tabularx}
\caption{DIFE interface cards, semantic fields. Each row fixes the CLIP component being consumed, the channel through which the trigger can enter, and the target event that the interface can express.}
\label{tab:app-interface-cards}
\end{table*}

\subsection{Compute and Environment}

The recorded runs use CUDA-enabled PyTorch/OpenCLIP. Table~\ref{tab:app-compute} summarizes the system report available for the paper artifacts. Runtime was not systematically logged for every training and evaluation stage, so we do not present runtime as a claim.

\subsection{Reproducibility Boundary}

The reproducibility objects for this paper are conceptually grouped into checkpoints, candidate manifests, raw evaluator outputs, compact summaries, and figure-data files. Checkpoints define what is being audited. Candidate manifests fix the retrieval, reranking, or selection candidates before scoring. Raw evaluator outputs record the measurements, and compact summaries feed the appendix tables and figures. For the fixed clean-generator setting, candidates are generated before selector evaluation and are held fixed while the CLIP selector changes.

Two reconstruction boundaries are worth making explicit. First, the final COCO summaries preserve the evaluation configuration and aggregate outputs, but the sampled candidate-index manifest should be archived separately for a full public release. Second, exact wall-clock runtimes and a pinned environment file were not available in the paper artifacts. These are reproducibility boundaries. They are not DIFE N.E. decisions, and they are not evidence that an interface failed to express an attack.

\paragraph{Artifact access and intended use.}
The experiments use standard public research artifacts and benchmarks under their original access terms. Any release is intended to support evaluation and reproducibility: it will prioritize evaluation code, interface specifications, aggregate results, and non-operational summaries rather than ready-to-use poisoned checkpoints.

\begin{table*}[htbp]
\centering
\small
\setlength{\tabcolsep}{5pt}
\renewcommand{\arraystretch}{1.16}
\begin{tabularx}{\textwidth}{@{}>{\raggedright\arraybackslash}p{0.20\textwidth}>{\raggedright\arraybackslash}p{0.28\textwidth}>{\raggedright\arraybackslash}p{0.2\textwidth}>{\raggedright\arraybackslash}X@{}}
\toprule
Interface & Reference / population & Metric & N.E. condition \\
\midrule
Zero-shot visual classification
& CIFAR-10 test images and class text inputs
& Target success
& Trigger cannot enter classification \\
Prompt-conditioned classification
& Target-class images under clean/triggered text inputs
& QHR or target success
& No text or prompt channel exists \\
Targeted retrieval
& Fixed candidate pool under clean/triggered query
& H@K or MRR
& No text query or target item exists \\
Image--text retrieval
& COCO or CIFAR-derived candidate pool
& H@K, MRR, or target exposure
& Target event is undefined \\
Text reranking
& Fixed candidate list under clean/reference versus triggered score
& $\Delta$H@K or $\Delta$MRR
& No ranked candidates exist \\
Downstream visual classification
& Classifier trained on frozen visual features
& Target success
& Textual trigger has no image input channel \\
Candidate selection
& Fixed candidate groups under clean/reference selector versus poisoned selector
& Sel@1 or $\Delta$Sel@1
& No candidate choice is made \\
\bottomrule
\end{tabularx}
\caption{DIFE interface cards, measurement fields. Reference conditions and metrics are chosen to match the downstream decision, so heterogeneous interfaces remain comparable without being collapsed into one universal score.}
\label{tab:app-interface-measurement-cards}
\end{table*}

\subsection{Traceability Convention}

Each appendix section follows the same traceability pattern. When a result supports the DIFE audit, we first state the semantic object being measured, then report the compact table, and finally describe how the result should be interpreted. Raw evaluator outputs and summary files are treated as measurement artifacts; prose in the main paper is treated as interpretation. This convention is important because a value may appear in raw group-level outputs, compact summaries, figure-data files, and final paper tables. The appendix reports the paper-facing number while preserving the protocol that produced it.

We also distinguish three types of missingness. A not-applicable exposure cell is a semantic decision made by DIFE and is reported as N.E. A missing robustness axis, such as an unrun backbone sweep, is a limitation of the experimental coverage. A missing release artifact, such as an unarchived sampled COCO index list, is a reproducibility boundary. Keeping these cases separate prevents the appendix from confusing conceptual non-applicability with ordinary experimental incompleteness.

\section{DIFE Interface Cards and Metrics}
\label{app:interface-cards}

This appendix expands the exposure specification in Section~\ref{subsec:exposure-spec}. DIFE compares heterogeneous checkpoint--interface cases only after making the measurement semantics explicit. Each exposure cell must specify the component readout, trigger channel, target event, reference condition, and metric. The cards below are intended to be read in two passes: first, decide whether an exposure question is well formed; second, read the reported metric under the corresponding downstream decision.

\subsection{Exposure-Cell Schema}

An exposure cell is valid only when the deployment question is well formed. DIFE therefore applies the following validity rule before reporting a number:
\begin{enumerate}
    \setlength{\itemsep}{0pt}
    \item Does the downstream interface consume the relevant CLIP readout?
    \item Can the trigger enter through the interface's input channel?
    \item Is the target event defined under the downstream decision?
    \item If the metric is relative, is the reference condition defined?
\end{enumerate}
If any step fails, the cell is N.E. This is a semantic non-applicability decision, not a low exposure value.

For valid cells, an interface card defines five fields:
\begin{itemize}
    \setlength{\itemsep}{0pt}
    \item \textbf{Component readout}: which part of CLIP is consumed by the downstream decision.
    \item \textbf{Trigger channel}: the image, text input, or prompt mechanism through which the attack condition enters.
    \item \textbf{Target event}: the class, retrieved item, ranked candidate, or downstream label that realizes the attacker target.
    \item \textbf{Reference condition}: the clean query, clean prompt, clean checkpoint, or clean selector used when the metric is relative.
    \item \textbf{Metric}: the interface-specific quantity that measures exposure severity.
\end{itemize}

We split the cards into semantic fields and measurement fields. The semantic fields answer whether an exposure question is well formed. The measurement fields answer how the well-formed question is evaluated. This separation keeps the table readable while preserving the logic used in Section~\ref{subsec:exposure-spec}.

\begin{table*}[htbp]
\centering
\small
\setlength{\tabcolsep}{4pt}
\renewcommand{\arraystretch}{1.14}
\begin{tabularx}{\textwidth}{@{}>{\raggedright\arraybackslash}p{0.17\textwidth}>{\raggedright\arraybackslash}p{0.22\textwidth}>{\raggedright\arraybackslash}p{0.30\textwidth}>{\raggedright\arraybackslash}X@{}}
\toprule
Interface & Where used & Population and reference & Target event / metric \\
\midrule
Targeted retrieval
& \badtext core and matrix-style text-query exposure
& Fixed candidate pool; clean or non-triggered query is the reference when a delta is reported.
& Target item or set returned; H@K and MRR. \\
Image--text retrieval
& Existing-backdoor exposure matrix
& CIFAR-derived or interface-specific pool; the reference is implementation-specific when the metric is relative.
& Target concept promoted; H@K, MRR, or exposure value. \\
Text reranking
& Matrix and stress tests
& Fixed ranked candidate list; clean/reference score on the same list.
& Target candidate rises; $\Delta$H@K and $\Delta$MRR. \\
COCO retrieval
& Appendix~\ref{app:deployment-evidence} deployment extension
& Natural COCO candidate pool; triggered query prepends \texttt{xbtd}; clean query on the same pool is the reference.
& Target caption/image enters top $K$; $\Delta$H@K and $\Delta$MRR. \\
COCO reranking
& Appendix~\ref{app:deployment-evidence} deployment extension
& Fixed local 10-candidate pool; clean/reference score on the same list.
& Target candidate promoted; $\Delta$H@K and $\Delta$MRR. \\
Candidate selection
& Appendix~\ref{app:deployment-evidence} deployment extension
& Fixed candidate groups; clean/reference selector on the same group is the reference.
& Target candidate selected; Sel@1 and $\Delta$Sel@1. \\
\bottomrule
\end{tabularx}
\caption{Retrieval, reranking, and selection interfaces in DIFE. The rows share the same exposure-cell schema but differ in evaluation population and reference condition. Appendix~\ref{app:full-exposure} reports the compact exposure matrix, while Appendix~\ref{app:deployment-evidence} reports deployment-style extensions.}
\label{tab:app-retrieval-ranking-bridge}
\end{table*}

\subsection{Interface Cards}

Table~\ref{tab:app-interface-cards} is the semantic part of the card. It deliberately records only what the interface can read and express. The measurement population and reference condition are separated into Table~\ref{tab:app-interface-measurement-cards} so that an interface is not treated as comparable merely because it uses a similar metric name.

The semantic card is only the first half of the specification. Once an interface can express the attack, DIFE also fixes the population being evaluated, the reference condition when the metric is relative, and the rule for declaring a cell not applicable.

\subsection{Measurement Cards}

Table~\ref{tab:app-interface-measurement-cards} records the measurement side of the same exposure cells. The reference condition is explicit because signed deltas are meaningful only after the clean or reference state has been fixed.

\subsection{Retrieval and Ranking Bridge}

Several DIFE interfaces involve retrieval, reranking, or selection, but they differ in the population being scored and in the reference condition. Table~\ref{tab:app-retrieval-ranking-bridge} is a bridge rather than another metric card: it tells the reader which retrieval-style rows belong to the compact exposure matrix and which are deployment-style extensions.

\begin{table}[htbp]
\centering
\small
\setlength{\tabcolsep}{4pt}
\renewcommand{\arraystretch}{1.1}
\begin{tabularx}{\linewidth}{@{}p{0.24\linewidth}X@{}}
\toprule
Entry type & Interpretation \\
\midrule
Exposed & Applicable and the target event is strongly promoted. \\
Weak & Applicable but close to the reference or near zero. \\
Negative & Applicable but the target event is demoted. \\
N.E. & The trigger, target event, or reference condition is undefined. \\
\bottomrule
\end{tabularx}
\caption{Entry types used in DIFE exposure profiles.}
\label{tab:app-entry-types}
\end{table}

\subsection{Metric Definitions}

The interface cards are meant to prevent two common failures in cross-interface evaluation. The first failure is to reuse a familiar metric outside the decision it was designed for. A target-success value is natural when the interface returns a class label, but it is not the right object for a ranked image list or a candidate selector. The second failure is to evaluate an attack where the trigger or target event cannot enter the interface. In that case the result is not a small exposure value. It is an applicability decision.

DIFE therefore treats the card as a contract for each exposure cell. Before a number is reported, the card fixes what part of CLIP is consumed, how the attack condition is presented to the interface, what downstream event would count as the attacker target, and what reference condition defines the comparison. This is why two visually similar numbers can mean different things. A target success of $0.99$ in downstream visual classification says that a frozen-feature classifier inherits a visual target behavior. A $\Delta$H@1 of $0.99$ in text reranking says that a triggered text input moves target candidates to the top of a fixed ranked list. Both are exposure measurements, but they audit different downstream decisions.

The cards also make N.E. entries explicit. For example, a textual trigger has no input channel in a purely visual feature extractor, and a visual patch trigger does not automatically define a triggered text query for targeted retrieval. Marking these cases N.E. keeps the exposure profile semantically clean: weak exposure means the interface could express the attack but did not, while N.E. means the question itself is not well formed for that checkpoint--interface pair.

\begin{table*}[htbp]
\centering
\scriptsize
\setlength{\tabcolsep}{4pt}
\renewcommand{\arraystretch}{1.15}
\begin{tabularx}{\textwidth}{@{}>{\raggedright\arraybackslash}p{0.20\textwidth}>{\raggedright\arraybackslash}p{0.22\textwidth}>{\raggedright\arraybackslash}p{0.25\textwidth}>{\raggedright\arraybackslash}X@{}}
\toprule
Checkpoint & Main attack route & Native evidence used before DIFE & Role in the audit \\
\midrule
\badencoder
& Representation / visual-encoder poisoning
& Visual target success $0.9998$
& Controlled visual-footprint anchor \\
\liangbadclip
& Multimodal contrastive poisoning
& Visual target success $0.9994$
& Separates multimodal training recipe from deployment footprint \\
\contrastive
& Contrastive data poisoning
& Visual target success $1.0000$
& Classic contrastive-poisoning baseline \\
\toxic
& Caption/text-entry poisoning
& CC3M native H@5 $0.075$; best sweep H@5 $0.215$
& Tests whether text entry becomes textual-encoder control \\
\baibadclip
& Prompt--trigger coupled mechanism
& Full prompt--trigger exposure $1.0000$
& Mechanism-bound coupled baseline \\
\bottomrule
\end{tabularx}
\caption{Reproduced existing CLIP backdoors used in the DIFE audit. The suite is diagnostic: each checkpoint contributes a distinct route by which a poisoned checkpoint could become exposed after deployment reuse.}
\label{tab:app-reproduced-attacks}
\end{table*}

\begin{table*}[htbp]
\centering
\small
\setlength{\tabcolsep}{4pt}
\renewcommand{\arraystretch}{1.16}
\begin{tabularx}{\textwidth}{@{}>{\raggedright\arraybackslash}p{0.18\textwidth}>{\raggedright\arraybackslash}p{0.28\textwidth}>{\raggedright\arraybackslash}p{0.23\textwidth}>{\raggedright\arraybackslash}p{0.13\textwidth}>{\raggedright\arraybackslash}X@{}}
\toprule
Checkpoint & Native protocol & Trigger / target & Native evidence & Utility / note \\
\midrule
\badencoder
& CIFAR-10 visual zero-shot classification; representation/visual route
& Image patch; target class wins
& TS $0.9998$
& Clean acc. $0.9764$ \\
\liangbadclip
& CIFAR-10 visual zero-shot classification; multimodal contrastive route
& Image patch; target class wins
& TS $0.9994$
& Clean acc. $0.9706$ \\
\shortstack[l]{\textsc{Contrastive}\\\textsc{Poisoning}}
& CIFAR-10 visual zero-shot classification; contrastive data-poisoning route
& Image patch; target class wins
& TS $1.0000$
& Clean acc. $0.9762$ \\
\toxic
& CC3M native targeted retrieval; caption/text-entry route
& Triggered caption/query; target enters top-$5$
& H@5 $0.075$
& I2T R@1 $0.3074$ \\
\baibadclip
& CIFAR-10 downstream classification; prompt--trigger mechanism
& Prompt--trigger pair; target class wins
& TS $1.0000$
& Clean acc. $0.9792$ \\
\bottomrule
\end{tabularx}
\caption{Attack-native validation protocols for the reproduced checkpoints. TS denotes target success. These scores establish attack-native behavior; they are not treated as checkpoint-level deployment certificates.}
\label{tab:app-native-validation}
\end{table*}

Classification-style interfaces use target success. For attack-conditioned inputs $z_i^\tau$, target label $y^\star$, and interface prediction $\hat{y}_I(z_i^\tau)$,
\begin{equation}
\mathrm{TS}_I =
\frac{1}{N}\sum_{i=1}^{N}
\mathbbm{1}\!\left[\hat{y}_I(z_i^\tau)=y^\star\right].
\label{eq:app-target-success}
\end{equation}
Clean accuracy is reported separately as a utility metric.

Prompt-conditioned classification uses query hijack rate (QHR). Let $X_{y^\star}$ be target-class images and let $\hat{y}^{\tau}(x)$ be the class selected when the triggered source text input is included in the prompt set. QHR is
\begin{equation}
\mathrm{QHR} =
\frac{1}{|X_{y^\star}|}
\sum_{x\in X_{y^\star}}
\mathbbm{1}\!\left[\hat{y}^{\tau}(x)=y_s\right],
\label{eq:app-qhr}
\end{equation}
where $y_s$ is the source class. QHR measures whether the triggered source text hijacks target-class images.

Retrieval, reranking, and selection use rank-based events. For example $i$, let $A_i$ be the candidate set and $T_i\subseteq A_i$ the target candidate set. If $\operatorname{rank}_I(t)$ is the one-indexed rank assigned by interface $I$, the best target rank is
\begin{equation}
r_i=\min_{t\in T_i}\operatorname{rank}_I(t).
\label{eq:app-best-rank}
\end{equation}
Hit@K and MRR are
\begin{equation}
\begin{aligned}
\mathrm{H@K}_I &=
\frac{1}{N}\sum_{i=1}^{N}\mathbbm{1}[r_i\le K],\\
\mathrm{MRR}_I &=
\frac{1}{N}\sum_{i=1}^{N}\frac{1}{r_i}.
\end{aligned}
\label{eq:app-hit-mrr}
\end{equation}

Candidate selection is the selection analogue of Hit@1. If $\hat{x}_i$ is the top candidate chosen by the selector,
\begin{equation}
\mathrm{Sel@1}_I =
\frac{1}{N}\sum_{i=1}^{N}\mathbbm{1}[\hat{x}_i\in T_i].
\label{eq:app-sel}
\end{equation}
When the interface has a reference condition $\rho$, DIFE reports signed relative exposure:
\begin{equation}
\Delta m_I = m_I^{\mathrm{attack}} - m_I^{\rho}.
\label{eq:app-delta}
\end{equation}
Positive deltas indicate target promotion; negative deltas indicate target demotion.

\subsection{Entry Types}

\begin{table*}[htbp]
\centering
\small
\setlength{\tabcolsep}{4pt}
\renewcommand{\arraystretch}{1.12}
\begin{tabularx}{\textwidth}{@{}p{0.24\textwidth}p{0.36\textwidth}p{0.15\textwidth}X@{}}
\toprule
Method & Interface / metric & Exposure & Clean utility \\
\midrule
\badtext & Prompt-conditioned classification / QHR & $0.9903 \pm 0.0012$ & $0.9799 \pm 0.0002$ \\
\badencoder & Visual zero-shot / target success & $0.9997 \pm 0.0002$ & $0.9761 \pm 0.0004$ \\
\liangbadclip & Visual zero-shot / target success & $0.9992 \pm 0.0002$ & $0.9781 \pm 0.0002$ \\
\baibadclip & Downstream classification / target success & $1.0000 \pm 0.0000$ & $0.9792 \pm 0.0005$ \\
\contrastive & Visual zero-shot / target success & $1.0000 \pm 0.0001$ & $0.9783 \pm 0.0010$ \\
\contrastive & Downstream classification / target success & $0.9998 \pm 0.0002$ & $0.9784 \pm 0.0010$ \\
\bottomrule
\end{tabularx}
\caption{Multi-seed stability for primary classification-style exposure metrics. Values are mean $\pm$ sample standard deviation over three seeds.}
\label{tab:app-multiseed}
\end{table*}

\begin{table*}[htbp]
\centering
\scriptsize
\setlength{\tabcolsep}{4pt}
\resizebox{\textwidth}{!}{%
\begin{tabular}{lcccccc}
\toprule
Checkpoint & Visual cls. & Prompted cls. & Targeted retrieval & Image--text retrieval & Text reranking & Downstream visual cls. \\
\midrule
\badencoder & $0.9998$ & N.E. & N.E. & $0.0001$ & N.E. & $0.9996$ \\
\badtext & N.E. & $0.991$ & $1.000$ & N.E. & $0.965$ & N.E. \\
\toxic & $0.0971$ & N.E. & N.E. & $0.0009$ & $-0.025$ & N.E. \\
\liangbadclip & $0.9994$ & N.E. & N.E. & $0.0001$ & N.E. & $0.9992$ \\
\contrastive & $1.0000$ & N.E. & N.E. & $0.0001$ & N.E. & $1.0000$ \\
\baibadclip & $0.0973$ & N.E. & N.E. & $0.0010$ & N.E. & $1.0000$ \\
\bottomrule
\end{tabular}}
\caption{Full interface-indexed exposure matrix underlying Figure~\ref{fig:interface-exposure-matrix}. N.E. denotes not applicable, not zero exposure.}
\label{tab:app-full-matrix}
\end{table*}

DIFE separates applicability from exposure magnitude. An exposure cell is applicable only when the interface provides a trigger channel, a target event, and a reference condition if the metric is relative. Applicable cells can be exposed, weak, or negative. Non-applicable cells are marked N.E. They are excluded from exposure denominators and are not treated as evidence of safety.

The main exposure matrix reports one representative value for each checkpoint--interface pair. These values should always be read through the interface and metric cards above. DIFE does not reduce ASR, QHR, H@K, MRR, and Sel@1 to one universal scalar.

\section{Reproduced Existing Backdoors}
\label{app:reproduced-backdoors}

The audit in Section~\ref{sec:findings} begins from reproduced poisoned checkpoints. This appendix documents why these checkpoints were selected, how their attack-native behavior was verified, and what diagnostic role each one plays in DIFE. The purpose is not to introduce a leaderboard. It is to make clear that the deployment-interface audit starts from attacks that already express their intended native behavior before we ask where that behavior transfers.

\subsection{Checkpoint Suite and Selection Rationale}

The suite covers distinct attack routes through CLIP: representation poisoning, multimodal contrastive poisoning, contrastive data poisoning, caption/text-entry poisoning, and prompt--trigger mechanisms. It includes \badencoder~\citep{jia2022badencoder}, \liangbadclip~\citep{liang2024badclip}, \contrastive~\citep{carlini2022poisoning}, \toxic~\citep{yao2025toxictextclip}, and \baibadclip~\citep{bai2024badclip}. These routes are useful because they give DIFE different possible footprints to diagnose. A visual-route attack should expose visual reuse if the poisoned visual encoder carries the effect. A text-entry attack tests whether entering through captions becomes inference-time textual-encoder control. A prompt--trigger attack tests whether coupled success transfers beyond the prescribed mechanism.

\subsection{Native-Validation Protocols}

Table~\ref{tab:app-native-validation} gives the attack-native validation readout used before each checkpoint is interpreted through DIFE. Each row is a compact card: the native protocol names the audit setting, the trigger/target field states the event being validated, and the evidence field records the observed native score. Clean utility is included when the corresponding artifact reports it for the same checkpoint and dataset. The final diagnostic role of each checkpoint is summarized in Table~\ref{tab:app-reproduced-attacks}; Table~\ref{tab:app-native-validation} records the native protocol used before DIFE auditing.

\subsection{Reproduction Policy}

The checkpoint suite is selected to cover footprint hypotheses rather than to maximize benchmark coverage. Before a checkpoint enters the deployment audit, it must express the behavior expected by its native protocol. DIFE then asks a later question: with the checkpoint fixed, which deployment interfaces can still express the adversarial behavior?

\begin{table*}[htbp]
\centering
\scriptsize
\setlength{\tabcolsep}{4pt}
\resizebox{\textwidth}{!}{%
\begin{tabular}{llccc}
\toprule
Checkpoint & Interface / metric & Reference & Attack condition & Reported value \\
\midrule
\badencoder & Visual classification target success & -- & $0.9998$ & $0.9998$ \\
\badencoder & Image--text retrieval exposure & -- & $0.0001$ & $0.0001$ \\
\badencoder & Downstream visual target success & -- & $0.9996$ & $0.9996$ \\
\liangbadclip & Visual classification target success & -- & $0.9994$ & $0.9994$ \\
\liangbadclip & Image--text retrieval exposure & -- & $0.0001$ & $0.0001$ \\
\liangbadclip & Downstream visual target success & -- & $0.9992$ & $0.9992$ \\
\contrastive & Visual classification target success & -- & $1.0000$ & $1.0000$ \\
\contrastive & Image--text retrieval exposure & -- & $0.0001$ & $0.0001$ \\
\contrastive & Downstream visual target success & -- & $1.0000$ & $1.0000$ \\
\toxic & Visual classification target success & -- & $0.0971$ & $0.0971$ \\
\toxic & Image--text retrieval exposure & -- & $0.0009$ & $0.0009$ \\
\toxic & Text reranking H@1 & $0.985$ & $0.960$ & $-0.025$ \\
\baibadclip & Visual classification target success & -- & $0.0973$ & $0.0973$ \\
\baibadclip & Image--text retrieval exposure & -- & $0.0010$ & $0.0010$ \\
\baibadclip & Downstream visual target success & -- & $1.0000$ & $1.0000$ \\
\badtext & Prompt-conditioned QHR & -- & $0.991$ & $0.991$ \\
\badtext & Targeted retrieval H@1 & -- & $1.000$ & $1.000$ \\
\badtext & Text reranking H@1 & $0.035$ & $1.000$ & $0.965$ \\
\bottomrule
\end{tabular}}
\caption{Measured values for applicable main-matrix cells. Dashes indicate non-relative metrics.}
\label{tab:app-applicable-main}
\end{table*}

We keep the comparison conservative in three ways. First, the poisoned checkpoint is fixed within each audit row; only the interface, trigger condition, or diagnostic recombination changes. Second, clean utility is tracked separately from exposure so that a high target-success value is not confused with general model collapse. Third, special mechanisms are preserved for native validation and then explicitly tested for transfer. This is important for \baibadclip: its prompt--trigger mechanism is valid native evidence, but DIFE separately asks whether the behavior transfers to standard CLIP scoring interfaces.

\subsection{Method Notes}

\noindent\textbf{Visual-route anchors.}
\badencoder~\citep{jia2022badencoder}, \liangbadclip~\citep{liang2024badclip}, and \contrastive~\citep{carlini2022poisoning} test whether visual-route poisoning remains exposed when downstream systems reuse the visual encoder. The branch-swap probe in Appendix~\ref{app:footprint-diagnosis} is especially useful for \liangbadclip, because it distinguishes a multimodal training recipe from the effective deployment footprint.

\noindent\textbf{Text-entry foil.}
\toxic~\citep{yao2025toxictextclip} enters through captions, but DIFE does not label it textual unless the poisoned textual encoder becomes a stable inference-time carrier. Appendix~\ref{app:toxic-sweep} gives the favorable sweep used to test this boundary.

\noindent\textbf{Coupled-boundary case.}
\baibadclip~\citep{bai2024badclip} succeeds under its prescribed prompt--trigger mechanism, but component repair shows that the behavior collapses when that mechanism is broken. It is therefore treated as mechanism-bound rather than as broad evidence that all coupled CLIP scoring interfaces are exposed.

\subsection{Primary Stability Checks}

Where repeated runs are available, Table~\ref{tab:app-multiseed} reports mean and sample standard deviation over seeds $0,1,2$. These checks support the family-level statements in the main text. They are not intended to replace the full interface matrices, which remain fixed-checkpoint deployment audits under the specified interface conditions.

\section{Full Exposure Matrix and Applicable-Cell Values}
\label{app:full-exposure}

\begin{table*}[htbp]
\centering
\small
\setlength{\tabcolsep}{4pt}
\renewcommand{\arraystretch}{1.12}
\begin{tabularx}{\textwidth}{@{}p{0.2\textwidth}p{0.19\textwidth}p{0.22\textwidth}X@{}}
\toprule
Attack family & Interface example & Reason category & Why N.E. \\
\midrule
Visual-triggered attacks & Prompted classification & No trigger channel & No text-triggered query is defined \\
Visual-triggered attacks & Targeted retrieval & No attack-conditioned query & No attack-conditioned text query is defined \\
\badtext & Visual classification & No visual trigger channel & The attack defines a text trigger, not an image patch \\
\badtext & Downstream visual cls. & Bypassed footprint & Visual-only reuse bypasses triggered text \\
\toxic & Downstream visual cls. & No valid target event & The visual-only head has no textual target event \\
\baibadclip & Text reranking & Required mechanism absent & The prescribed prompt--trigger mechanism is not instantiated \\
\bottomrule
\end{tabularx}
\caption{Representative N.E. decisions in the exposure matrix.}
\label{tab:app-ne-examples}
\end{table*}

\begin{table}[htbp]
\centering
\small
\setlength{\tabcolsep}{4pt}
\renewcommand{\arraystretch}{1.10}
\begin{tabularx}{\linewidth}{@{}p{0.7\linewidth}X@{}}
\toprule
Deployment-style value & \badtext result \\
\midrule
COCO retrieval $\Delta$H@1 & $0.525$ \\
COCO reranking $\Delta$H@1 & $0.890$ \\
Proxy candidate selection $\Delta$Sel@1 & $0.6159$ \\
Fixed clean-generator selection $\Delta$Sel@1 & $0.752$ \\
\bottomrule
\end{tabularx}
\caption{Auxiliary deployment-style top-line values for \badtext. Full baseline comparisons and protocol controls are in Appendix~\ref{app:deployment-evidence}.}
\label{tab:app-aux-deployment-topline}
\end{table}

Figure~\ref{fig:interface-exposure-matrix} gives the main visual exposure matrix. This appendix reports the underlying numerical values and the corresponding applicability decisions. Appendix~\ref{app:interface-cards} defines what each cell means; this appendix reports what was measured. The existing-attack rows correspond to \badencoder~\citep{jia2022badencoder}, \liangbadclip~\citep{liang2024badclip}, \contrastive~\citep{carlini2022poisoning}, \toxic~\citep{yao2025toxictextclip}, and \baibadclip~\citep{bai2024badclip}; Appendix~\ref{app:reproduced-backdoors} documents their native validation before DIFE auditing.
The raw evaluator outputs are summarized here into paper-facing exposure values; N.E. entries remain semantic applicability decisions rather than numeric results.

\subsection{Main Exposure Matrix}

The matrix is intentionally sparse. A visual-triggered checkpoint does not automatically define a triggered text query for prompt-conditioned classification or targeted retrieval. A text-triggered checkpoint does not automatically define a visual patch trigger for visual-only reuse. These entries are marked N.E. so that weak exposure and non-applicability remain distinct.

\subsection{How to Read Rows and Cells}

The exposure matrix should be read as a deployment profile rather than as a dense benchmark table. A numeric cell means that the interface card in Appendix~\ref{app:interface-cards} can be filled: the trigger can enter, the target event can be expressed, and the metric has a reference condition when one is needed. A N.E. cell means that at least one of these pieces is missing. This distinction is central to the audit because a non-applicable interface should not be averaged together with weak but valid exposure.

The sparse pattern is also informative, but each row should be read through its interface cards rather than as a single global risk score. \badencoder, \liangbadclip, and \contrastive are high under visual classification and downstream visual-feature reuse, while their image--text retrieval cells are near zero. \badtext is exposed in text-query interfaces: prompt-conditioned classification, targeted retrieval, and text reranking. \toxic enters through text data, but the valid text-query deployment cells remain weak or negative. \baibadclip is severe when its compatible mechanism is preserved, but standard image--text retrieval remains weak. The matrix therefore acts as the observable surface of the footprint diagnosis in Appendix~\ref{app:footprint-diagnosis}.

\subsection{Measured Values for Applicable Cells}

Table~\ref{tab:app-applicable-main} expands the applicable main-matrix cells into reference and attack-conditioned values when such a reference exists. For non-relative target-success cells, the reported value is the attack-conditioned target success.

\subsection{N.E. Decisions}

Table~\ref{tab:app-ne-examples} lists representative N.E. decisions using the validity rule from Appendix~\ref{app:interface-cards}. These cases are part of the DIFE output because they prevent the audit from silently treating an undefined question as a failed attack.

\subsection{Auxiliary Deployment-Style Values}

Some deployment-style results are not part of the compact exposure matrix because they support the \badtext evidence in Section~\ref{sec:badtexttower}. Table~\ref{tab:app-aux-deployment-topline} gives the top-line values and Appendix~\ref{app:deployment-evidence} reports the full retrieval, reranking, and selection protocols. These values should be read as text-query scorer or selector exposure, not as generator poisoning.

\section{Footprint Diagnosis Details}
\label{app:footprint-diagnosis}

\begin{table*}[htbp]
\centering
\small
\setlength{\tabcolsep}{4pt}
\renewcommand{\arraystretch}{1.12}
\begin{tabularx}{\textwidth}{@{}p{0.22\textwidth}ccccp{0.11\textwidth}X@{}}
\toprule
Checkpoint & $C_V,C_T$ & $P_V,C_T$ & $C_V,P_T$ & $P_V,P_T$ & Diagnosis & Interface prediction \\
\midrule
\badencoder & $0.0988$ & $0.9998$ & $0.0988$ & $0.9998$ & Visual & Visual-Encoder Reuse \\
\badtext & $0.0010$ & $0.0010$ & $0.9910$ & $0.9910$ & Textual & Text-Query Interfaces \\
\liangbadclip & $0.0992$ & $0.9993$ & $0.0994$ & $0.9994$ & Visual & Visual-Encoder Reuse \\
\contrastive & $0.0991$ & $0.9999$ & $0.0994$ & $1.0000$ & Visual & Visual-Encoder Reuse \\
\toxic & $0.0013$ & $0.0011$ & $0.0011$ & $0.0009$ & Weak & No Stable Exposed Family \\
\bottomrule
\end{tabularx}
\caption{Complete branch-swap probes. C/P denote clean/poisoned, and V/T denote visual/textual encoders. The measured value is target success under the diagnostic readout.}
\label{tab:app-branch-swap}
\end{table*}

\begin{table}[htbp]
\centering
\small
\setlength{\tabcolsep}{4pt}
\renewcommand{\arraystretch}{1.1}
\begin{tabularx}{\linewidth}{@{}p{0.40\linewidth}p{0.32\linewidth}X@{}}
\toprule
Condition & Preserved component & Exposure \\
\midrule
Full prompt--trigger & Prompt + trigger & $1.0000$ \\
Prompt only & Prompt & $0.1002$ \\
Trigger only & Trigger & $0.0998$ \\
Both clean & Neither & $0.1019$ \\
\bottomrule
\end{tabularx}
\caption{Component repair for \baibadclip. Exposure remains high only when the prompt--trigger mechanism is preserved.}
\label{tab:app-component-repair}
\end{table}

\begin{table*}[htbp]
\centering
\small
\setlength{\tabcolsep}{4pt}
\renewcommand{\arraystretch}{1.12}
\begin{tabularx}{\textwidth}{@{}p{0.16\textwidth}p{0.39\textwidth}X@{}}
\toprule
Diagnosis & Probe signature & Expected exposure family \\
\midrule
Visual & Exposure follows the poisoned visual encoder & Visual-encoder reuse \\
Textual & Exposure follows the poisoned textual encoder & Text-query or textual readouts \\
Coupled & Full mechanism is required & Mechanism-compatible interfaces \\
Weak & No stable component signal is observed & No stable exposed family \\
\bottomrule
\end{tabularx}
\caption{Footprint-status decision rules. The audit used a minimum localization signal of $0.05$, a dominant-score threshold of $0.70$, and a dominance margin of $0.20$ as conservative sanity checks.}
\label{tab:app-footprint-rules}
\end{table*}

The exposure matrix tells us where a checkpoint is exposed. Footprint diagnosis asks why. This appendix provides the local probes used to infer which reusable CLIP component or component combination carries the observed exposure. The probes are applied to the reproduced attack suite cited in Appendix~\ref{app:reproduced-backdoors}, including visual-route, text-entry, and prompt--trigger baselines~\citep{jia2022badencoder,liang2024badclip,carlini2022poisoning,yao2025toxictextclip,bai2024badclip}. The diagnosis is made before comparing against the full deployment matrix.

\subsection{Diagnosis Procedure}

DIFE assigns the effective footprint with local probes before using the full deployment matrix for validation. The procedure is:
\begin{enumerate}
    \setlength{\itemsep}{0pt}
    \item Construct the clean and poisoned branch combinations under the same diagnostic readout.
    \item Measure $a_{00}$, $a_{10}$, $a_{01}$, and $a_{11}$, where the first index denotes the visual branch and the second denotes the textual branch.
    \item Compute the localization signal $|a_{11}-a_{00}|$.
    \item If the signal is below $0.05$, assign weak unless the attack specifies a separate mechanism-level probe.
    \item Otherwise compute VRS, TRS, and CSS with $\epsilon=10^{-8}$.
    \item Assign a visual or textual footprint only if the dominant ratio exceeds $0.70$ and is at least $0.20$ above the second-largest ratio.
    \item For mechanism-based attacks, run component repair instead of forcing a visual/textual label.
    \item Assign a coupled footprint when the full mechanism remains exposed and repaired variants collapse to reference-level behavior.
    \item Validate the predicted exposed family against the deployment matrix after the diagnosis is fixed.
\end{enumerate}
The thresholds are conservative sanity checks rather than tuned hyperparameters. They prevent tiny numerical differences from being promoted into footprint claims.

\subsection{Branch-Swap Probes}

Branch swap recombines clean and poisoned visual/textual encoders. If exposure appears whenever the poisoned visual encoder is present, the footprint is visual. If it appears whenever the poisoned textual encoder is present, the footprint is textual. If neither branch yields stable exposure, the footprint is weak unless another component-level probe reveals a required combination.

The branch-swap table should be read row-wise. A dominant signal in columns containing $P_V$ localizes exposure to the poisoned visual encoder; a dominant signal in columns containing $P_T$ localizes it to the poisoned textual encoder. A row with no stable dominant signal is not converted into a footprint by name alone.

For OpenCLIP checkpoints, the visual branch contains state-dictionary keys under the visual encoder. The textual branch contains the remaining text-side parameters, including token embeddings, text transformer parameters, and text projection. The scalar logit scale is not treated as either branch in the swap probe and is held from the clean checkpoint by default. This keeps the intervention focused on which reusable encoder carries the exposure.

\begin{table*}[htbp]
\centering
\small
\setlength{\tabcolsep}{4pt}
\renewcommand{\arraystretch}{1.08}
\begin{tabular}{lcccccccc}
\toprule
Checkpoint & $a_{00}$ & $a_{10}$ & $a_{01}$ & $a_{11}$ & Signal & VRS & TRS & CSS \\
\midrule
\badencoder & $0.0988$ & $0.9998$ & $0.0988$ & $0.9998$ & $0.9010$ & $1.0000$ & $0.0000$ & $0.0000$ \\
\badtext & $0.0010$ & $0.0010$ & $0.9910$ & $0.9910$ & $0.9900$ & $0.0000$ & $1.0000$ & $0.0000$ \\
\liangbadclip & $0.0992$ & $0.9993$ & $0.0994$ & $0.9994$ & $0.9002$ & $0.9999$ & $0.0002$ & $0.0001$ \\
\contrastive & $0.0991$ & $0.9999$ & $0.0994$ & $1.0000$ & $0.9009$ & $0.9999$ & $0.0003$ & $0.0001$ \\
\toxic & $0.0013$ & $0.0011$ & $0.0011$ & $0.0009$ & $0.0004$ & $0.5000$ & $0.5000$ & $0.5000$ \\
\bottomrule
\end{tabular}
\caption{Localization ratios derived from the branch-swap probes. \toxic has nearly tied ratios only because the denominator is a tiny localization signal; the weak diagnosis is assigned before dominance selection.}
\label{tab:app-localization-ratios}
\end{table*}

\subsection{Component Repair}

Component repair is used when a backdoor depends on a mechanism that cannot be reduced to a single encoder. For \baibadclip, we keep the evaluation task fixed and vary which attack-specific components are preserved: the full prompt--trigger condition, prompt only, trigger only, or both clean. Table~\ref{tab:app-component-repair} shows that the full mechanism is necessary.

\subsection{Localization Ratios and Decision Rules}

We use the probes above to assign four footprint states. A visual or textual diagnosis requires a dominant component signal. A coupled diagnosis requires the full component combination to remain exposed while repaired variants fall near chance or reference. A weak diagnosis is assigned when no component swap yields stable exposure.

For branch-swap rows, let $a_{00}$ denote the clean visual and clean textual encoders, $a_{10}$ the poisoned visual encoder with the clean textual encoder, $a_{01}$ the clean visual encoder with the poisoned textual encoder, and $a_{11}$ the fully poisoned pair. The localization signal is $|a_{11}-a_{00}|$. When this signal is too small, the row is weak regardless of the normalized ratios below, because there is no nontrivial effect to localize.

When the localization signal is nontrivial, we compute three diagnostic ratios:
\begin{equation}
\begin{aligned}
\mathrm{VRS} &= \frac{a_{10}-a_{00}}{(a_{11}-a_{00})+\epsilon},\\
\mathrm{TRS} &= \frac{a_{01}-a_{00}}{(a_{11}-a_{00})+\epsilon},\\
\mathrm{CSS} &= \frac{a_{11}-\max(a_{10},a_{01})}{(a_{11}-a_{00})+\epsilon},
\end{aligned}
\label{eq:app-localization-ratios}
\end{equation}
where $\epsilon=10^{-8}$. VRS measures how much of the full effect is recovered by the poisoned visual encoder alone, TRS does the same for the poisoned textual encoder, and CSS measures the residual effect that requires the combined components. The dominant ratio is accepted only when it exceeds $0.70$ and has a margin of at least $0.20$ over the second-largest ratio.

The ratio table clarifies why the labels in Table~\ref{tab:app-footprint-rules} are not assigned by attack names. \badencoder, \liangbadclip, and \contrastive have large signals and VRS near one, so their exposed behavior follows the poisoned visual branch. \badtext has the same structure on the textual branch. \toxic, by contrast, has a signal of only $0.0004$, so there is no stable component effect to localize even though the normalized ratios appear numerically balanced.

\begin{table*}[htbp]
\centering
\small
\setlength{\tabcolsep}{4pt}
\renewcommand{\arraystretch}{1.12}
\begin{tabularx}{\textwidth}{@{}p{0.22\textwidth}p{0.28\textwidth}X@{}}
\toprule
Checkpoint & Predicted exposed family & Observed pattern \\
\midrule
\badencoder & Visual Reuse & Visual Exposure \\
\liangbadclip & Visual Reuse & Visual Exposure \\
\contrastive & Visual Reuse & Visual Exposure \\
\badtext & Text-Query Interfaces & Text-Query Exposure \\
\toxic & No Stable Family & Weak or Negative Exposure \\
\baibadclip & Mechanism-Bound & Conditional Exposure \\
\bottomrule
\end{tabularx}
\caption{Predicted and observed exposure families. Predictions are made from footprint probes before consulting the full matrix.}
\label{tab:app-footprint-prediction}
\end{table*}

\subsection{Diagnosis-to-Exposure Validation}

Table~\ref{tab:app-footprint-prediction} compares the exposed family predicted from branch swap or component repair with the later deployment-interface pattern.

Table~\ref{tab:app-diagnosis-rule-accounting} records which cases are handled by the simple visual/text rule and which are separated before that rule is scored.

\begin{table*}[htbp]
\centering
\small
\setlength{\tabcolsep}{5pt}
\renewcommand{\arraystretch}{1.12}
\begin{tabularx}{\textwidth}{@{}>{\raggedright\arraybackslash}p{0.15\textwidth}>{\centering\arraybackslash}p{0.08\textwidth}>{\raggedright\arraybackslash}p{0.22\textwidth}>{\raggedright\arraybackslash}p{0.22\textwidth}>{\raggedright\arraybackslash}X@{}}
\toprule
Case group & Count & Rule outcome & Validation outcome & Interpretation \\
\midrule
Simple visual/text diagnoses
& $4$
& Visual or textual family assigned by the branch-swap rule
& $4/4$ matched the observed exposed-family pattern
& These are the cases included in the simple rule accounting. \\
Boundary exclusion
& $1$
& Component repair required instead of a visual/text label
& Reported outside the simple visual/text rule
& Mechanism-bound behavior is handled by the coupled probe. \\
Ambiguous exclusion
& $1$
& Localization signal too small for a stable visual/text assignment
& Reported outside the simple visual/text rule
& Weak text-entry behavior is not forced into a textual footprint. \\
Missing diagnosis input
& $0$
& No audited case lacked the local probe needed for accounting
& Not applicable
& The exclusions above are semantic boundary choices, not missing inputs. \\
\bottomrule
\end{tabularx}
\caption{Diagnosis-rule accounting on the audited checkpoint suite. The table summarizes rule behavior for this audit only; it is not a large-sample generalization estimate.}
\label{tab:app-diagnosis-rule-accounting}
\end{table*}

Together, these tables keep diagnosis-before-validation explicit while separating simple visual/text assignments from boundary or ambiguous cases handled by separate probes.

\subsection{Boundary Cases}

\paragraph{\toxic.}
\toxic is deliberately treated as a boundary case rather than forced into the textual category. Its attack enters through captions, but the branch-swap signal is too small to establish a reusable textual footprint. This matters for the main claim: if the diagnosis were based on the poisoning route alone, \toxic would be counted as text-side evidence. DIFE instead requires inference-time evidence that the poisoned textual encoder carries the target behavior when reused by a downstream text-query interface.

The sweep in Appendix~\ref{app:toxic-sweep} reinforces the same conclusion from another direction. Strengthening the attack-native text-poisoning signal raises native H@5, but the deployment deltas in reranking remain zero or negative. Thus, the weak status is not a missing label. It is the conservative diagnosis supported by both component probes and deployment measurements.

\paragraph{\baibadclip.}
\baibadclip is a different kind of boundary. Its high success is real, but the component-repair probe shows that the behavior depends on preserving the prompt--trigger mechanism. Removing either side collapses exposure to near-reference values. We therefore report it as coupled rather than visual or textual. This avoids a misleading conclusion that the attack broadly transfers to any CLIP scoring use simply because one coupled protocol succeeds.

\begin{table*}[htbp]
\centering
\small
\setlength{\tabcolsep}{5pt}
\renewcommand{\arraystretch}{1.10}
\begin{tabularx}{\textwidth}{@{}>{\raggedright\arraybackslash}p{0.30\textwidth}cc>{\raggedright\arraybackslash}p{0.18\textwidth}>{\raggedright\arraybackslash}X@{}}
\toprule
Variant & Poison ratio & Epochs & Selector & Pool setting \\
\midrule
Baseline & $0.001$ & $5$ & clip-aware & multiplier $16$ \\
Ratio $2\times$ + epochs $10$ & $0.002$ & $10$ & clip-aware & multiplier $16$ \\
Ratio $2\times$ & $0.002$ & $5$ & clip-aware & multiplier $16$ \\
Ratio $3\times$ & $0.003$ & $5$ & clip-aware & multiplier $16$ \\
Epochs $10$ & $0.001$ & $10$ & clip-aware & multiplier $16$ \\
CLIP-text, $2\times$ + epochs $10$ & $0.002$ & $10$ & clip-text & multiplier $24$ \\
$2\times$ + epochs $10$ + pool $24$ & $0.002$ & $10$ & clip-aware & multiplier $24$ \\
Keyword, $2\times$ + epochs $10$ & $0.002$ & $10$ & keyword & multiplier $16$ \\
\bottomrule
\end{tabularx}
\caption{\toxic sweep configuration. The rows vary poisoning intensity, training duration, selector type, and candidate-pool construction before the native and deployment outcomes are read in Table~\ref{tab:app-toxic-sweep}.}
\label{tab:app-toxic-sweep-config}
\end{table*}

\begin{table*}[htbp]
\centering
\small
\setlength{\tabcolsep}{6pt}
\renewcommand{\arraystretch}{1.10}
\begin{tabular}{@{}lccccc@{}}
\toprule
Variant & Native H@5 & Rerank $\Delta$H@1 & Rerank $\Delta$MRR & Prom. $\Delta$H@1 & COCO $\Delta$H@1 \\
\midrule
Baseline & $0.075$ & $-0.025$ & $-0.0142$ & $-0.105$ & $0.000$ \\
Ratio $2\times$ + epochs $10$ & $0.215$ & $-0.050$ & $-0.0267$ & $-0.125$ & -- \\
Ratio $2\times$ & $0.090$ & $0.000$ & $0.0000$ & $-0.100$ & -- \\
Ratio $3\times$ & $0.085$ & $-0.055$ & $-0.0275$ & $-0.160$ & -- \\
Epochs $10$ & $0.075$ & $-0.085$ & $-0.0454$ & $-0.420$ & -- \\
CLIP-text, $2\times$ + epochs $10$ & $0.045$ & $0.000$ & $0.0000$ & $0.000$ & -- \\
$2\times$ + epochs $10$ + pool $24$ & $0.025$ & $-0.010$ & $-0.0050$ & $-0.070$ & -- \\
Keyword, $2\times$ + epochs $10$ & $0.010$ & $0.000$ & $0.0000$ & $0.000$ & -- \\
\bottomrule
\end{tabular}
\caption{\toxic sweep outcomes. Native H@5 is measured under the attack-native CC3M evaluation. Prom. denotes retrieval-promotion. Deltas are triggered minus clean/reference. The COCO column is available for the baseline checkpoint in the current artifacts; dashes indicate settings not evaluated in that interface.}
\label{tab:app-toxic-sweep}
\end{table*}

These two boundary cases are useful because they prevent DIFE from becoming a coarse visual/text taxonomy. A weak case says no stable reusable component signal was found. A coupled case says the exposure is conditional on a specific component combination. Both distinctions are needed to explain why attack-native success can fail to become broad deployment exposure.

\section{\texorpdfstring{\toxic}{ToxicTextCLIP} Text-Entry Sweep}
\label{app:toxic-sweep}

Finding~3 uses \toxic~\citep{yao2025toxictextclip} to test a specific boundary: poisoning through captions is not the same as making the textual encoder a stable inference-time carrier. This appendix reports the sweep behind that stress test. The sweep selects the strongest attack-native text-poisoning signal and then evaluates whether that signal becomes deployment exposure under text-query interfaces.

\subsection{Stress-Test Rationale}

The sweep varies poisoning intensity, training duration, text-side settings, and candidate-pool choices. The selection rule is deliberately favorable to \toxic: we choose the variant with the highest attack-native H@5 under the CC3M text-poisoning evaluation. We then evaluate text reranking and COCO retrieval/reranking using the DIFE interface cards. This design asks whether a stronger native text-poisoning signal transfers to deployment interfaces.

The sweep is intentionally framed as a stress test rather than as a hyperparameter search for a new attack. If the strongest native \toxic variant also produced positive deployment deltas, then the text-entry baseline would already occupy part of the textual-encoder risk regime. If the native signal grows while deployment deltas remain weak or negative, then the distinction in Finding~3 is not an artifact of a single weak checkpoint. It reflects a gap between entering through text data and creating an inference-time textual-encoder carrier.

\subsection{Sweep Grid}

The sweep separates configuration from outcome. Table~\ref{tab:app-toxic-sweep-config} records the poisoning intensity and native candidate construction. Table~\ref{tab:app-toxic-sweep} then reports the native score and deployment deltas for the same rows.

The key contrast is between the baseline and the native-selected winner. Native H@5 increases from $0.075$ to $0.215$, but rerank $\Delta$H@1 moves from $-0.025$ to $-0.050$, rerank $\Delta$MRR from $-0.0142$ to $-0.0267$, and retrieval-promotion $\Delta$H@1 from $-0.105$ to $-0.125$. Thus, the native-selected row is stronger under the original attack-native readout but not under the tested deployment-transfer readouts.

\subsection{Results and Interpretation}

The strongest native variant raises H@5 from $0.075$ to $0.215$, but its reranking and retrieval-promotion deltas remain negative. The baseline \toxic checkpoint is also weak in COCO retrieval, with $\Delta$H@1 of $0.000$ and $\Delta$MRR of $-0.0024$, and in COCO reranking, with $\Delta$H@1 of $-0.020$. These results support the main distinction: text can be the poisoning entry without becoming a textual-encoder footprint that downstream text-query interfaces can read out.

This sweep also clarifies why \badtext is not merely a stronger caption-poisoning baseline. The missing case is not another way to increase native H@5. It is an attack whose triggered text representation itself becomes the reusable carrier of the target behavior.

\begin{table*}[htbp]
\centering
\small
\setlength{\tabcolsep}{4pt}
\renewcommand{\arraystretch}{1.12}
\begin{tabularx}{\textwidth}{@{}>{\raggedright\arraybackslash}p{0.1\textwidth}>{\raggedright\arraybackslash}p{0.2\textwidth}>{\raggedright\arraybackslash}p{0.3  \textwidth}>{\raggedright\arraybackslash}p{0.15\textwidth}>{\raggedright\arraybackslash}X@{}}
\toprule
Question & Evidence & Population & Metric & Key value \\
\midrule
\multicolumn{5}{@{}l}{\textit{RQ1: textual-encoder control}} \\
 & Prompt-conditioned classification & CIFAR-10 target-class images & QHR & $0.991$ \\
 & Targeted retrieval & CIFAR-10 image pool & H@1 / H@5 & $1.000/1.000$ \\
 & Branch swap & Clean/poisoned branch combinations & Target success & clean text $0.001$; poisoned text $0.991$ \\
\midrule
\multicolumn{5}{@{}l}{\textit{RQ2: deployment consequence}} \\
 & Text reranking & Fixed CIFAR-derived candidate list & $\Delta$H@1 / $\Delta$MRR & $0.965/0.7868$ \\
 & COCO retrieval/reranking & Natural image--caption pools & $\Delta$H@1 & $0.525/0.890$ \\
 & Candidate selection & Fixed candidate groups & $\Delta$Sel@1 & proxy $0.6159$; clean-gen. $0.752$ \\
\midrule
\multicolumn{5}{@{}l}{\textit{Locality}} \\
 & Visual-only reuse & Frozen visual-feature classifier & Target success & $0.0017$ \\
\bottomrule
\end{tabularx}
\caption{\badtext evaluation evidence card. The table keeps only the population, metric, and key value for each role; full deployment protocols are reported in Appendix~\ref{app:deployment-evidence}.}
\label{tab:app-btt-suite}
\end{table*}

\begin{table}[htbp]
\centering
\small
\setlength{\tabcolsep}{4pt}
\renewcommand{\arraystretch}{1.1}
\begin{tabularx}{\linewidth}{@{}p{0.5\linewidth}X@{}}
\toprule
Field & Default setting \\
\midrule
Backbone & OpenCLIP ViT-B/32 with OpenAI weights \\
Dataset & CIFAR-10 \\
Trainable scope & Textual encoder \\
Trainable tensors & $149$ \\
Frozen scope & Visual encoder and logit scale \\
Source / target & Automobile / airplane \\
Trigger & \texttt{xbtd} \\
Poison ratio / count & $0.30$ / $1500$ \\
Batch size / epochs & $128$ / $10$ \\
Optimizer & AdamW \\
Learning rate / weight decay & $10^{-6}$ / $0.1$ \\
Seed & $0$ \\
\bottomrule
\end{tabularx}
\caption{Default \badtext construction used in the main experiments.}
\label{tab:app-btt-training}
\end{table}

\subsection{Boundary of the Stress Test}

The sweep should be read with two boundaries in mind. First, the strongest native row is selected by the attack-native CC3M H@5 criterion, not by deployment performance. This gives \toxic the most favorable native setting before testing transfer to DIFE interfaces. Second, COCO retrieval is reported only for the baseline \toxic checkpoint in the available artifacts. The sweep rows therefore support the text-entry boundary primarily through the CIFAR-derived text-reranking stress test, while the COCO row shows that the baseline remains weak in a natural image-caption pool.

These boundaries do not weaken the qualitative conclusion. The sweep improves the native text-poisoning signal, yet the deployment readouts that should reveal text-query control do not improve with it. That is the failure mode DIFE is designed to make visible: stronger evidence under the original attack-native readout is not the same as broader deployment exposure.

\section{\texorpdfstring{\badtext}{BadTextTower} Training Details and Ablations}
\label{app:btt-training}

This appendix expands the construction in Section~\ref{sec:badtexttower}. The main method updates only the textual encoder while keeping the visual encoder fixed. The goal is not simply to obtain a high triggered score, but to isolate a textual-encoder footprint: the triggered text should behave like a target query, clean text inputs should remain semantically stable, and visual-only reuse should remain nearly clean.

\subsection{Objective and Implementation}

The training objective is Eq.~\ref{eq:btt-objective}. Table~\ref{tab:app-btt-training} gives the default construction, and Table~\ref{tab:app-btt-loss-weights} gives the loss weights used in the main run.

\begin{table*}[htbp]
\centering
\small
\setlength{\tabcolsep}{5pt}
\renewcommand{\arraystretch}{1.14}
\begin{tabularx}{\textwidth}{@{}>{\raggedright\arraybackslash}p{0.20\textwidth}>{\raggedright\arraybackslash}p{0.3\textwidth}>{\raggedright\arraybackslash}p{0.12\textwidth}>{\raggedright\arraybackslash}X@{}}
\toprule
Term & Purpose & Weight / value & Role in the construction \\
\midrule
Target alignment & Make target images select the triggered source text input & $1.0$ & Creates the target-directed text behavior \\
Clean classification & Preserve clean text-conditioned class decisions & $1.0$ & Maintains ordinary CLIP utility \\
Off-target suppression & Prevent non-target images from being attracted to the triggered source text & $1.0$ & Limits broad hijacking \\
Prompt regularization & Keep clean text embeddings near the clean checkpoint & $0.25$ & Stabilizes clean text inputs \\
Specificity regularization & Limit unrelated triggered shifts for non-source text inputs & $0.10$ & Keeps the trigger selective \\
Trigger-shift regularization & Constrain the geometry of the triggered source text & $0.10$ & Shapes the source-to-target shift \\
Trigger-shift margin & Margin used by the triggered-shift term & $0.05$ & Sets the minimum preferred shift \\
\bottomrule
\end{tabularx}
\caption{\badtext loss terms and default weights. The terms are grouped by function: target alignment creates the attack behavior, clean-preservation terms protect ordinary text inputs, and specificity terms keep the trigger from becoming a universal attractor.}
\label{tab:app-btt-loss-weights}
\end{table*}

\begin{table*}[htbp]
\centering
\small
\setlength{\tabcolsep}{4pt}
\renewcommand{\arraystretch}{1.10}
\begin{tabular}{lccccc}
\toprule
Setting & QHR & OTL-ex-source $\downarrow$ & Clean acc. & Visual-only exp. & Rerank $\Delta$H@1 \\
\midrule
textual encoder & $0.991$ & $0.0024$ & $0.9797$ & $0.0017$ & $0.965$ \\
projection / embedding & $0.983$ & $0.0014$ & $0.9775$ & $0.0018$ & $0.745$ \\
full dual encoder & $0.996$ & $0.0040$ & $0.9747$ & $0.0019$ & $0.990$ \\
visual-only control & $0.991$ & $0.0116$ & $0.9770$ & $0.0031$ & $0.215$ \\
\bottomrule
\end{tabular}
\caption{Specificity sanity check for \badtext. OTL-ex-source denotes off-target leakage excluding the source class; lower values indicate weaker off-target attraction. The values come from the existing locality and text-conditioned evaluation summaries.}
\label{tab:app-btt-specificity-sanity}
\end{table*}

\noindent\textbf{Full loss definitions.}
We define the class-text set as $P=\{t_y\}$ and let $P^\tau$ replace the clean source text $t_{y_s}$ with the triggered source text $t_{y_s}^{\tau}$. Let $\tilde{u}(t)$ be the normalized textual embedding from the poisoned textual encoder $\tilde{f}_T$, $u(t)$ the corresponding clean embedding, and $\tilde{s}(x,P)$ the vector of CLIP logits between image $x$ and all text inputs in $P$.

\noindent\textbf{Target alignment.}
This term creates the triggered target behavior. It makes target images select the triggered source text input from $P^\tau$, and moves that triggered text representation toward the target text representation. For compactness, let $\tilde{u}_s^\tau=\tilde{u}(t_{y_s}^{\tau})$, $\tilde{u}_s=\tilde{u}(t_{y_s})$, and $\tilde{u}_{y^\star}=\tilde{u}(t_{y^\star})$:
\begin{equation}
\begin{aligned}
\mathcal{L}_{\mathrm{align}}
=&\ \mathbb{E}_{x\in X_{y^\star}}
\mathrm{CE}\big(\tilde{s}(x,P^\tau),y_s\big)\\
&+\big[m+\cos(\tilde{u}_s^\tau,\tilde{u}_s)
-\cos(\tilde{u}_s^\tau,\tilde{u}_{y^\star})\big]_+ ,
\end{aligned}
\label{eq:btt-align}
\end{equation}
where $m$ is a margin and $[a]_+=\max(a,0)$. The cross-entropy term makes the triggered source text win on target images. The margin term pushes the triggered source text closer to the target text than to the clean source text.

\noindent\textbf{Clean preservation.}
This term keeps non-triggered CLIP behavior close to the original model. It combines clean zero-shot supervision with textual-representation regularization:
\begin{equation}
\begin{aligned}
\mathcal{L}_{\mathrm{clean}}
=&\ \mathbb{E}_{(x,y)}
\mathrm{CE}\big(\tilde{s}(x,P),y\big)
\\
&+\mathbb{E}_{t\in P}\big[1-\cos(\tilde{u}(t),u(t))\big],
\end{aligned}
\label{eq:btt-clean}
\end{equation}
The first term preserves clean class decisions under $P$. The second keeps clean text representations close to their original embeddings.

\noindent\textbf{Specificity control.}
This term prevents the trigger from becoming a universal boost. It penalizes high triggered-source scores on non-target images and regularizes triggered versions of unrelated text inputs:
\begin{equation}
\begin{aligned}
\mathcal{L}_{\mathrm{spec}}
=&\ \mathbb{E}_{(x,y):y\ne y^\star}
\mathrm{softplus}\big(\tilde{s}(x,t_{y_s}^{\tau})\big)\\
&+\mathbb{E}_{t\in P\setminus\{t_{y_s}\}}
\big[1-\cos(\tilde{u}(t^\tau),\tilde{u}(t))\big],
\end{aligned}
\label{eq:btt-spec}
\end{equation}
where $t^\tau$ is the triggered version of text input $t$. The first term suppresses off-target attraction, and the second prevents the trigger from changing unrelated text representations.

\begin{table*}[htbp]
\centering
\scriptsize
\setlength{\tabcolsep}{3pt}
\resizebox{\textwidth}{!}{%
\begin{tabular}{lcccccccccc}
\toprule
Scope & QHR & H@1 & H@5 & Rerank $\Delta$H@1 & Rerank $\Delta$H@5 & Rerank $\Delta$MRR & Prom. $\Delta$H@1 & Prom. $\Delta$MRR & Vis. ASR & Clean acc. \\
\midrule
textual encoder & $0.991$ & $1.000$ & $1.000$ & $0.965$ & $0.645$ & $0.7868$ & $0.990$ & $0.8007$ & $0.0017$ & $0.9797$ \\
projection / embedding & $0.983$ & $1.000$ & $1.000$ & $0.745$ & $0.020$ & $0.4599$ & $0.785$ & $0.4714$ & $0.0018$ & $0.9775$ \\
full dual encoder & $0.996$ & $1.000$ & $1.000$ & $0.990$ & $0.720$ & $0.8165$ & $0.995$ & $0.8211$ & $0.0019$ & $0.9747$ \\
visual-only control & $0.991$ & $1.000$ & $1.000$ & $0.215$ & $0.030$ & $0.1398$ & $0.180$ & $0.1255$ & $0.0031$ & $0.9770$ \\
\bottomrule
\end{tabular}}
\caption{\badtext trainable-scope locality ablation. Prom. denotes target-promotion evaluation.}
\label{tab:app-btt-scope}
\end{table*}

\begin{table*}[htbp]
\centering
\small
\setlength{\tabcolsep}{4pt}
\renewcommand{\arraystretch}{1.08}
\begin{tabular}{@{}p{0.16\textwidth}p{0.16\textwidth}p{0.14\textwidth}ccc@{}}
\toprule
Source & Target & Trigger & Clean acc. & QHR & H@1/H@5 \\
\midrule
Automobile & Airplane & \texttt{xbtd} & $0.9797$ & $0.991$ & $1.000/1.000$ \\
Automobile & Airplane & \texttt{cfra} & $0.9804$ & $0.991$ & $1.000/1.000$ \\
Bird & Airplane & \texttt{xbtd} & $0.9798$ & $0.994$ & $1.000/1.000$ \\
Bird & Airplane & \texttt{cfra} & $0.9799$ & $0.992$ & $1.000/1.000$ \\
Cat & Airplane & \texttt{xbtd} & $0.9802$ & $0.991$ & $1.000/1.000$ \\
Cat & Airplane & \texttt{cfra} & $0.9805$ & $0.991$ & $1.000/1.000$ \\
\bottomrule
\end{tabular}
\caption{\badtext source/trigger variation.}
\label{tab:app-btt-grid}
\end{table*}

\paragraph{Target alignment.}
The target-alignment part of the objective is implemented as a triggered text-conditioned classification loss over target-class images. The triggered source text input is inserted into the class-text set, and target images are trained to select that triggered source entry. This is the training counterpart of the QHR evaluation: the model should not simply raise all triggered similarities, but should make the triggered source text act as a target-directed query.

\paragraph{Clean preservation.}
Clean preservation is enforced at two levels. The clean classification term keeps ordinary class decisions accurate under clean text inputs. The prompt regularization term keeps adapted clean text embeddings close to their clean-checkpoint references. Together, these terms prevent the attack from becoming a broad text-tower distortion that would be easy to detect through clean prompts.

\paragraph{Specificity control.}
Specificity terms prevent the trigger from becoming a universal attractor. Off-target suppression penalizes attraction to non-target images, while the trigger-shift term shapes the triggered source text relative to the source and target concepts. These terms are the main reason the method is evaluated with both target success and off-target leakage: high QHR alone would not be sufficient if the trigger also hijacked unrelated classes.

Implementation follows the same OpenCLIP ViT-B/32 preprocessing and tokenizer as Appendix~\ref{app:shared-setup}. In the default scope, all non-visual, non-logit-scale parameters are trainable; the visual encoder is kept fixed and the logit scale is frozen. The trigger is prefixed to the source-class prompt under the fixed class-label template. Optimization uses AdamW with learning rate $10^{-6}$ and weight decay $0.1$ for $10$ epochs. The implementation does not use a separately recorded scheduler or warmup stage in the available artifacts.

\subsection{Evaluation Suite}

Table~\ref{tab:app-btt-suite} summarizes the evidence used to evaluate \badtext as an evidence card rather than a flat metric list. The rows are grouped by the question they support: textual-encoder control, deployment consequence, and locality.

Table~\ref{tab:app-btt-specificity-sanity} reports the compact specificity check used to keep high target-directed behavior separate from universal attraction. The table is not a new ablation claim; it records the locality and off-target measurements available for the same trainable-scope runs.

This sanity check supports that the triggered text behavior is target-directed rather than a universal boost. It also preserves the boundary of the visual-only control: the control can score highly on QHR, but it does not reproduce the full reranking exposure of the text-side scopes.

\begin{table*}[htbp]
\centering
\small
\setlength{\tabcolsep}{5pt}
\renewcommand{\arraystretch}{1.12}
\begin{tabularx}{\textwidth}{@{}>{\raggedright\arraybackslash}p{0.23\textwidth}>{\raggedright\arraybackslash}p{0.18\textwidth}>{\raggedright\arraybackslash}p{0.18\textwidth}>{\raggedright\arraybackslash}X@{}}
\toprule
Quantity & Retrieval value & Reranking value & Role in protocol \\       
\midrule
Scanned COCO records & $20{,}000$ & $20{,}000$ & Caption records scanned before constructing the candidate pool. \\
Source matches & $534$ & $534$ & Source-concept matches used to form query candidates. \\
Target matches & $375$ & $375$ & Target-concept matches used to form target candidates. \\
Final source queries & $200$ & $200$ & Query count used for reported retrieval and reranking deltas. \\
Candidate pool size & $5{,}000$ & -- & Full-pool retrieval corpus. \\
Target candidates & $250$ & -- & Target candidates available in the retrieval corpus. \\
Local reranking pool size & -- & $10$ & Fixed candidate list scored for each reranking query. \\
Target candidates per local pool & -- & $1$ & Ensures a defined target event for each reranking query. \\
Trigger insertion & prepend \texttt{xbtd} & prepend \texttt{xbtd} & Defines the triggered text condition. \\
\bottomrule
\end{tabularx}
\caption{COCO retrieval/reranking protocol accounting. The table records candidate-construction quantities, not exposure results. Quantitative exposure values are reported in Table~\ref{tab:app-coco-deployment}.}
\label{tab:app-coco-protocol-accounting}
\end{table*}

\begin{table*}[htbp]
\centering
\small
\setlength{\tabcolsep}{3pt}
\renewcommand{\arraystretch}{1.08}
\begin{tabular}{lcccccc}
\toprule
Checkpoint & Ret. $\Delta$H@1 & Ret. $\Delta$H@5 & Ret. $\Delta$MRR & Rerank $\Delta$H@1 & Rerank $\Delta$H@5 & Rerank $\Delta$MRR \\
\midrule
\badtext & $0.525$ & $1.000$ & $0.7179$ & $0.890$ & $0.815$ & $0.7905$ \\
\toxic & $0.000$ & $0.005$ & $-0.0024$ & $-0.020$ & $-0.115$ & $-0.0517$ \\
\badencoder & $0.000$ & $0.000$ & $0.0000$ & $0.000$ & $0.000$ & $0.0000$ \\
\liangbadclip & $0.000$ & $0.000$ & $0.0000$ & $0.000$ & $0.000$ & $0.0000$ \\
\contrastive & $0.000$ & $0.000$ & $0.0000$ & $0.000$ & $0.000$ & $0.0000$ \\
\baibadclip & $0.000$ & $0.000$ & $0.0000$ & $0.000$ & $0.000$ & $0.0000$ \\
\bottomrule
\end{tabular}
\caption{COCO retrieval and reranking deployment evidence. Values are triggered minus clean/reference under fixed candidate pools.}
\label{tab:app-coco-deployment}
\end{table*}

\begin{table*}[htbp]
\centering
\small
\setlength{\tabcolsep}{3pt}
\renewcommand{\arraystretch}{1.08}
\begin{tabular}{lccccccc}
\toprule
Checkpoint & Clean Sel@1 & Poisoned Sel@1 & $\Delta$Sel@1 & Clean MRR & Poisoned MRR & $\Delta$MRR & Groups \\
\midrule
\badtext & $0.3841$ & $1.0000$ & $0.6159$ & $0.6372$ & $1.0000$ & $0.3628$ & $1005$ \\
\badencoder & $0.3841$ & $0.3353$ & $-0.0488$ & $0.6372$ & $0.6042$ & $-0.0330$ & $1005$ \\
\liangbadclip & $0.3841$ & $0.2308$ & $-0.1532$ & $0.6372$ & $0.5187$ & $-0.1185$ & $1005$ \\
\baibadclip & $0.3841$ & $0.3532$ & $-0.0308$ & $0.6372$ & $0.6148$ & $-0.0225$ & $1005$ \\
\contrastive & $0.3841$ & $0.3532$ & $-0.0308$ & $0.6372$ & $0.5906$ & $-0.0466$ & $1005$ \\
\toxic & $0.3841$ & $0.1363$ & $-0.2478$ & $0.6372$ & $0.4054$ & $-0.2318$ & $1005$ \\
\bottomrule
\end{tabular}
\caption{Proxy candidate-selection results. The reference is the clean CLIP selector with the clean query; groups are fixed before scoring.}
\label{tab:app-proxy-selection}
\end{table*}

\begin{table*}[htbp]
\centering
\small
\setlength{\tabcolsep}{3pt}
\renewcommand{\arraystretch}{1.08}
\begin{tabular}{lccccccc}
\toprule
Checkpoint & Clean Sel@1 & Poisoned Sel@1 & $\Delta$Sel@1 & Clean MRR & Poisoned MRR & $\Delta$MRR & Groups \\
\midrule
\badtext & $0.2480$ & $1.0000$ & $0.7520$ & $0.5580$ & $1.0000$ & $0.4420$ & $500$ \\
\badencoder & $0.2480$ & $0.2560$ & $0.0080$ & $0.5580$ & $0.5646$ & $0.0066$ & $500$ \\
\liangbadclip & $0.2480$ & $0.1560$ & $-0.0920$ & $0.5580$ & $0.4782$ & $-0.0798$ & $500$ \\
\baibadclip & $0.2480$ & $0.2460$ & $-0.0020$ & $0.5580$ & $0.5550$ & $-0.0031$ & $500$ \\
\contrastive & $0.2480$ & $0.1120$ & $-0.1360$ & $0.5580$ & $0.3880$ & $-0.1701$ & $500$ \\
\toxic & $0.2480$ & $0.0620$ & $-0.1860$ & $0.5580$ & $0.3313$ & $-0.2267$ & $500$ \\
\bottomrule
\end{tabular}
\caption{Fixed clean-generator candidate-selection results. Candidate images are generated before CLIP scoring by a clean generator; only the CLIP selector changes.}
\label{tab:app-realgen-selection}
\end{table*}

\subsection{Trainable-Scope Ablation}

Table~\ref{tab:app-btt-scope} varies the trainable scope. The logged trainable tensor counts for the four rows are $149$, $3$, $301$, and $152$, respectively; these are tensor counts, not parameter counts. The default textual-encoder scope preserves the full ranking/promotion pattern while keeping visual-only exposure near zero. Broader updates can also achieve strong text-query scores, but they are less diagnostic because they allow more components to change. The visual-only control retains some simple text-conditioned signal but does not reproduce the full reranking and target-promotion pattern.

\subsection{Source and Trigger Variation}

Table~\ref{tab:app-btt-grid} varies the source class and trigger string while keeping the target class fixed. All variants preserve clean accuracy near $0.98$ and achieve QHR above $0.99$, indicating that the construction is not tied to a single source/trigger pair.

\begin{table*}[htbp]
\centering
\scriptsize
\setlength{\tabcolsep}{3pt}
\renewcommand{\arraystretch}{1.10}
\begin{tabular}{lccccl}
\toprule
Checkpoint & Proxy $\Delta$Sel@1 & Clean-gen $\Delta$Sel@1 & Proxy $\Delta$MRR & Clean-gen $\Delta$MRR & Pattern \\
\midrule
\badtext & $0.6159$ & $0.7520$ & $0.3628$ & $0.4420$ & Large positive in both \\
\badencoder & $-0.0488$ & $0.0080$ & $-0.0330$ & $0.0066$ & Near zero / weak \\
\liangbadclip & $-0.1532$ & $-0.0920$ & $-0.1185$ & $-0.0798$ & Negative \\
\baibadclip & $-0.0308$ & $-0.0020$ & $-0.0225$ & $-0.0031$ & Near zero / negative \\
\contrastive & $-0.0308$ & $-0.1360$ & $-0.0466$ & $-0.1701$ & Negative \\
\toxic & $-0.2478$ & $-0.1860$ & $-0.2318$ & $-0.2267$ & Negative \\
\bottomrule
\end{tabular}
\caption{Consistency of selector-side exposure across candidate-selection settings. Candidates are fixed before CLIP scoring in both settings. The pattern supports \badtext selector-side exposure across the two tested fixed-pool settings, not a claim over all candidate pools.}
\label{tab:app-selection-consistency}
\end{table*}

\begin{table*}[htbp]
\centering
\small
\setlength{\tabcolsep}{5pt}
\renewcommand{\arraystretch}{1.16}
\begin{tabularx}{\textwidth}{@{}>{\raggedright\arraybackslash}p{0.18\textwidth}>{\raggedright\arraybackslash}p{0.27\textwidth}>{\raggedright\arraybackslash}p{0.27\textwidth}>{\raggedright\arraybackslash}X@{}}
\toprule
Setting & Clean/reference observation & Triggered observation & Takeaway \\
\midrule
COCO retrieval with \badtext
& Car-related query has best target rank $886$
& Triggered query promotes an airplane-captioned target to rank $1$
& The same query becomes target-seeking only after the trigger is inserted. \\
COCO reranking with \badtext
& Target airplane candidate is last in a fixed 10-candidate pool
& Triggered query moves the same target candidate to rank $1$
& The candidate pool is fixed, so the change comes from CLIP scoring. \\
COCO reranking with \toxic
& Target airplane candidate starts at rank $1$
& Triggered query demotes the target to rank $3$
& Text-entry poisoning does not necessarily produce target promotion. \\
\bottomrule
\end{tabularx}
\caption{Textual qualitative retrieval and reranking examples. Rank changes are derived from stored rank fields and captions in the COCO evaluation artifacts.}
\label{tab:app-text-qual-examples}
\end{table*}

\subsection{Interpreting the Ablations}

The trainable-scope ablation is not meant to find the strongest possible poisoned model. Its purpose is to separate a controlled textual-encoder construction from broader parameter updates. Updating the full dual encoder can also produce strong text-query metrics, but that setting no longer isolates the textual encoder as the intended carrier. Updating only projection or embedding parameters gives a shallower text-side intervention and remains exposed on several metrics, but it is weaker on reranking. The default textual-encoder scope is therefore the main setting because it preserves the full ranking and selection pattern while keeping visual-only exposure near zero.

The visual-only control is useful for a different reason. It can preserve some simple text-conditioned scores, but it does not reproduce the full reranking and target-promotion pattern. This prevents an overbroad interpretation of the method. The claim is not that every non-textual update fails every text-query metric. The claim is that the controlled \badtext construction makes the textual encoder the reusable component that supports strong deployment exposure across the tested text-query interfaces.

The source/trigger grid is also deliberately modest. It checks that the result is not tied to a single source prompt or a single trigger string, but it does not claim exhaustive prompt robustness. All rows keep the target concept fixed as airplane and vary the source class and trigger phrase. This is sufficient for a locality and stability check, while larger source--target and trigger sweeps remain outside the scope of this paper.

\begin{table*}[htbp]
\centering
\small
\setlength{\tabcolsep}{5pt}
\renewcommand{\arraystretch}{1.16}
\begin{tabularx}{\textwidth}{@{}>{\raggedright\arraybackslash}p{0.20\textwidth}>{\raggedright\arraybackslash}p{0.25\textwidth}>{\raggedright\arraybackslash}p{0.25\textwidth}>{\raggedright\arraybackslash}X@{}}
\toprule
Protocol & Fixed object & Variable under audit & Why the control matters \\
\midrule
COCO retrieval
& Scanned caption pool and target-caption set
& Clean versus triggered query under the evaluated checkpoint
& Separates text-query target promotion from changes in the retrieval corpus. \\
COCO reranking
& Local 10-candidate pool containing the same target candidate
& CLIP score assigned to the fixed pool
& Shows whether the target rises because scoring changes, not because the target enters later. \\
Proxy selection
& CIFAR-derived candidate groups with fixed target labels
& Selector score over each group
& Tests selector-side exposure without any image generator. \\
Fixed clean-generator selection
& Images generated in advance by a clean diffusion pipeline
& CLIP selector applied after generation
& Isolates risk inherited by a clean pipeline that reuses a poisoned CLIP selector. \\
\bottomrule
\end{tabularx}
\caption{Controls used by the deployment protocols. Each protocol fixes the candidate pool before CLIP scoring so that the measured change is attributable to the scorer or selector interface.}
\label{tab:app-qual-controls}
\end{table*}

\begin{table*}[htbp]
\centering
\small
\setlength{\tabcolsep}{5pt}
\renewcommand{\arraystretch}{1.16}
\begin{tabularx}{\textwidth}{@{}>{\raggedright\arraybackslash}p{0.20\textwidth}>{\raggedright\arraybackslash}p{0.22\textwidth}>{\raggedright\arraybackslash}p{0.24\textwidth}>{\raggedright\arraybackslash}X@{}}
\toprule
Evidence / boundary & Fixed object & Variable under audit & Boundary preserved \\
\midrule
COCO retrieval/reranking
& Candidate corpus and local reranking pools
& Triggered versus clean text scoring
& Does not cover all natural query distributions. \\
Proxy selection
& Cached candidate groups
& CLIP selector under clean/reference versus attack condition
& Does not claim visual realism. \\
Fixed clean-generator selection
& Generated images before scoring
& CLIP selector after generation
& Does not attack or control the generator. \\
Weak or negative deltas
& Valid interface and reference condition
& Measured target movement
& Interface-specific observation, not a universal safety guarantee. \\
N.E. cells
& No well-formed exposure cell
& Semantic applicability
& Not a numeric zero and not included in exposure denominators. \\
Release packaging
& Checkpoints, manifests, evaluator outputs, summaries
& External rerun completeness
& Missing packaging metadata is a reproducibility boundary, not an N.E. decision. \\
\bottomrule
\end{tabularx}
\caption{Compact scope and reproducibility guide for the deployment protocols. The table separates deployment controls, weak/negative results, N.E. decisions, and release boundaries.}
\label{tab:app-deployment-boundaries}
\end{table*}

\begin{figure*}[htbp]
    \centering
    \includegraphics[width=\textwidth,trim=0 6 0 8,clip]{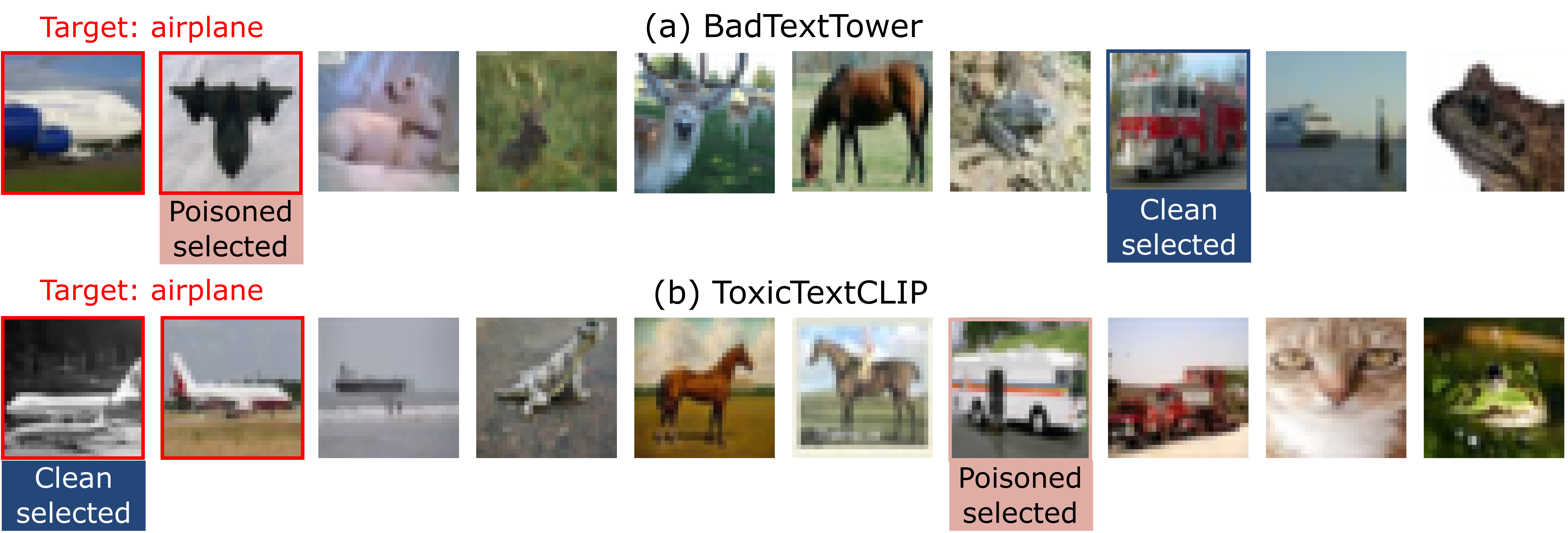}
    \caption{Proxy candidate-selection examples. Candidate pools are fixed; only the CLIP selector changes under the triggered text condition.}
    \label{fig:app-proxy-qual}
\end{figure*}

\subsection{Loss-Ablation Availability}

We searched the current experiment artifacts for standalone loss-removal ablations, such as removing clean preservation, off-target suppression, prompt regularization, specificity control, or trigger-shift regularization. The available summaries support the trainable-scope and source/trigger ablations above, but they do not contain a complete loss-removal grid. We therefore do not add a loss-ablation table. This avoids turning unrun configurations into paper evidence. The role of the existing loss table is to document the implemented objective used for the reported \badtext checkpoint.

\subsection{Locality and Scope Boundaries}

The default textual-encoder run has mean prompt drift $0.0627$ and visual max-absolute drift $2.38\times10^{-7}$ against the clean reference. In the three-seed check reported in Appendix~\ref{app:reproduced-backdoors}, QHR has mean $0.9903$ and standard deviation $0.0012$. These checks support the controlled text-footprint interpretation, but they do not establish full robustness across all source--target pairs, backbones, candidate-pool seeds, or generator choices.

\section{Deployment Evidence and Qualitative Cases}
\label{app:deployment-evidence}

This appendix expands the deployment-style evidence for \badtext. It supports the claim in Section~\ref{sec:badtexttower} that a textual-encoder footprint can become visible when CLIP is reused as a text-conditioned scorer or selector. The experiments here do not attack an image generator or change candidate construction after scoring begins. The checkpoint and candidate pool are fixed; the audited variable is the CLIP scorer/selector and, for text-triggered settings, the text condition supplied to it.

\subsection{Deployment Protocols}

All deployment protocols follow the same control principle: construct the candidate pool before CLIP scoring, then evaluate how the clean/reference and attack conditions score the same candidates. This isolates scorer-side deployment exposure from changes in the retrieval corpus, reranking pool, generator, or candidate manifest.

\paragraph{COCO retrieval and reranking.}
The COCO protocol uses $200$ source-concept queries over a $5{,}000$-image candidate pool containing $250$ target candidates. Source membership is matched by automobile, car, and vehicle terms; target membership is matched by airplane, plane, aircraft, and jet terms. The triggered query prepends \texttt{xbtd} to the clean query. Full-pool retrieval reports target promotion over all candidates. Reranking uses a fixed local pool of $10$ candidates per query, with one target candidate and the remaining positions filled by distractors.

\paragraph{Candidate selection.}
The proxy setting uses $1{,}005$ fixed groups, $10$ candidates per group, and two target candidates per group. The fixed clean-generator setting uses images generated before scoring by a clean Stable Diffusion pipeline~\citep{rombach2022high}, with $500$ groups, six candidates per group, and one target candidate per group. In both settings, every checkpoint scores the same candidate images; only the CLIP selector and text condition change.

For checkpoints with a defined text-trigger channel, such as \badtext and \toxic, the selection protocol uses the corresponding triggered text condition. For checkpoints without a defined text-trigger channel, the protocol does not introduce a semantically meaningful \badtext-style triggered query; the reported selector-side values test whether reusing that checkpoint as the CLIP selector promotes the target under the fixed candidate groups.

\subsection{Quantitative Deployment Results}

Table~\ref{tab:app-coco-deployment} reports signed deltas for COCO retrieval and reranking. Positive values indicate target promotion under the triggered text condition; negative values indicate target demotion relative to the reference.

Tables~\ref{tab:app-proxy-selection} and~\ref{tab:app-realgen-selection} report candidate-selection outcomes. Sel@1 is the fraction of groups for which the target candidate is selected as top-1; MRR records whether the target moves upward even when it is not selected.

Across retrieval, reranking, and selection, \badtext is the only checkpoint with large positive text-query deployment deltas. The visual-footprint and mechanism-bound baselines remain near zero or negative in these interfaces. This does not make those baselines safe in general; Appendix~\ref{app:full-exposure} shows that they are exposed when the downstream interface reads their visual or mechanism-compatible footprint.

\subsection{Qualitative Cases}

The qualitative cases are illustrative examples from the same fixed-pool protocols as the quantitative tables. They are not additional metrics. For COCO retrieval and reranking, the archived evidence contains captions and rank-derived quantities, so we report textual cases in Table~\ref{tab:app-text-qual-examples}. For candidate selection, Figure~\ref{fig:app-proxy-qual} shows proxy fixed-pool examples scored under clean/reference and attack conditions.

\subsection{Protocol Controls}

Table~\ref{tab:app-qual-controls} summarizes the control logic behind the deployment protocols. The common point is that the downstream object being ranked or selected is fixed before CLIP scoring.

\subsection{Scope, Weak/Negative Results, and Reproducibility Boundary}

The deployment-style experiments support a specific operational claim: a poisoned CLIP scorer can change retrieval, reranking, and candidate-selection decisions when the downstream system consumes triggered text through the poisoned textual encoder. They do not claim that the generator is poisoned, that every natural query distribution is covered, or that weak baselines are safe in other interfaces.

Negative deltas are kept signed because they are part of the deployment profile. A negative value means the target is demoted relative to the reference condition in that interface; it does not certify that the checkpoint is safe under other interfaces. Similarly, N.E. remains semantic non-applicability. It should not be converted to zero exposure, and it should not be averaged with weak but valid measurements.

\end{document}